\documentclass[preprint,12pt]{elsarticle}
\usepackage{graphicx}
\usepackage{float}
\usepackage{amsthm}
\theoremstyle{plain}
\newtheorem*{theorem*}{Theorem}

\usepackage{amssymb}
\usepackage{amsthm}
\usepackage{amsmath}
\usepackage{graphics}
\usepackage{pstricks}
\DeclareFixedFont{\fiverm}{OT1}{crm}{m}{n}{5pt}
\input{prepictex}
\input{pictex}
\input{postpictex}
\usepackage{stackrel}
\theoremstyle{plain}  
\theoremstyle{definition} 
\newtheorem*{defn*}{Definition.} 
\theoremstyle{remark}

\begin{document}

\begin{frontmatter}

\title{Non-Perturbative Quantum Correction to the Reissner-Nordstr\"om spacetime with Running Newton's Constant}

\address[UNorte]{Departamento de F\'{i}sica y Geociencias, Universidad del Norte. Km 5 via a Puerto Colombia AA 1569}
\author[UNorte]{O. Ruiz}
\author[UNorte]{E. Tuiran}

\begin{abstract}
We study the consequences of the running Newton's constant on several key aspects of spherically symmetric charged black holes by performing a renormalization group improvement of the classical Reissner-Nordstr\"om metric within the framework of the Einstein-Hilbert truncation in quantum Einstein gravity. In particular, we determine that the event horizon surfaces are stable except for the extremal case and we corroborate the appearance of a new extremality condition at the Planck scale that hints at the existence of a final stage after the black hole evaporation process. We find explicit expressions for the area and surface gravity of the event horizon and we show the existence of an exact form $dS$ with the surface gravity as integrating factor. This result is a first clue towards finding a generalization of the first law of black hole dynamics, for improved charged spherically symmetric spacetimes in contrast to a previous no-go result for axially symmetric ones with null charge. We finally calculate an explicit formula for the Komar mass at the event horizon, and we compare it with the total mass at infinity for a wide domain of values of the black hole parameters $M$, $Q$ and $\bar{w}$, showing a loss of mass which can be interpreted as a consequence of the antiscreening effect of the gravitational field between the event horizon and infinity.

\end{abstract}


\end{frontmatter}
\section{Introduction}
\label{S:1}

During the past decades many efforts have been drawn to the exploration of the non-perturbative behavior of quantum gravity. \cite{Reuter_96}-\cite{Percacci}. In reference \cite{Reuter_96} M. Reuter introduced  an exact renormalization group (RG) functional equation for quantum gravity; that defines the evolution under renormalization group transformations  of coupling constants in the space of theories. This equation has been applied in the frame of the so-called Einstein-Hilbert truncation; which permits a non-perturbative approximation to the RG-flow of the Newton and cosmological constants \cite{Reuter_96}. The complete flow pattern for this truncation has been derived \cite{Saueressig}, and finer truncations have been studied \cite{Lauscher, Lauscher1, Codello}. The influence of matter fields has been taken into account, \cite{Dou, Perini}, and the RG flow has been optimized in \cite{Litim} based on the results from \cite{Reuter_96} and \cite{Lauscher}. The most important consequence of these investigations is the prediction of the existence of a non-Gaussian fixed point (NGFP) for the RG flow of quantum gravity \cite{Souma}. After detailed studies of the reliability of the pertinent truncations \cite{Saueressig, Lauscher, Lauscher1, Litim} it is currently believed the existence of a fixed point in the exact theory that projects itself at every stage of truncation and is by no means an artifact of any approximation. It has been found that it posesses all the necessary properties to render quantum gravity a non-perturbatively renormalizable theory, solving in this way the difficulties that arise from an approach with perturbation theory. The line of research in quantum gravity where the existence of the mentioned non-gaussian fixed point plays a fundamental role, is called the asymptotic safety scenario \cite{Weinberg, Niedermaier, L_Reuter}. 

The exact renormalization group flow equation (ERGE) is a theoretical tool that provides a non-perturbative approach to the task of finding quantum corrections to the exact solutions of the Einstein field equations. It is based on the hypothesis that general theory of relativity is an effective theory at low energies coming from a fundamental one, valid at higher energy scales \cite{Reuter_96}. The term Quantum Einstein Gravity (QEG) is refered to this fundamental theory and its consequences.

Concerning finding non-perturbative quantum corrections to Einstein gravity, three possible levels of improvement have been so far explored \cite{Bonanno-Reuter,R-Tuiran, Saueressig-1,Saueressig-2,Tuiran,BR_IR,Weyer1,Weyer2}. A first level implies substituting the relevant running couplings directly in the metric components; at the second level the substitution takes place in the field equations; the third level is the highest in deductive power, namely substituting the running couplings directly in the action. In this work we study exclusively the influence of the running Newton constant and we carry out the improvement at the lowest first level. The consequences of this improvement are modest in the sense of their range of application, but the procedure is simpler in its execution, and the previous results we already have at disposal in the literature in order to compare with the present work, are implemented at the first level.

Several of the previous results concerning quantum improvement of classical black hole space-times make contact, directly or indirectly with the Reissner-Nordstr\"{o}m solution \cite{Bonanno-Reuter,R-Tuiran,Tuiran,Koch-Gonz,Ishibashi}; we are specially interested in the following ones \cite{Bonanno-Reuter,R-Tuiran,Tuiran}: 

\begin {enumerate}[i]
\item Several expressions derived for the improved Schwarzschild black hole notoriously resemble the same expressions for the classical Reissner-Nordstr\"{o}m spacetime.

\item The original form of the first law of black hole mechanics for the improved Kerr spacetime doesn't hold anymore, and the relation between Hawking temperature and surface gravity is no longer directly proportional. 

\item The calculation of the Komar integral for the mass of the improved Schwarzschild and Kerr black holes indicates an effect of antiscreening due to the gravitational quantum field.    
\end{enumerate}
The above mentioned results serve as a starting point for the present work. 

\noindent \\ Reissner-Nordstr\"{o}m's solution is a spherically symmetric spacetime represented by 
\begin{eqnarray}
ds^{2}=-f(r)dt^{2}+f(r)^{-1}dr^{2}+r^{2}d\Omega^{2}, \label{RN_1} \\
f(r)\equiv 1-\frac{2G_0 M}{r}+\frac{G_0 Q^{2}}{r^{2}},
\label{Class RN}
\end{eqnarray}
\noindent where $d\Omega ^{2}\equiv d\theta ^{2}+\sin ^{2}\theta d\phi ^{2}$
is the infinitesimal element of squared solid angle, $G_0$ is the experimentally observed current value of the Newton's constant and $Q$ and $M$ are the charge and mass of the black hole respectively.
Assuming as a main hypothesis that beyond the classical limit, for sufficiently long distances, the principal effects from quantum gravity related to linear fluctuations of the metric, are included in the dependence of the Newton constant on the energy scale, or RG infrared cutoff $k$ \cite{Reuter_1}; and after identifying the cutoff $k$ with the inverse of a scalar proper distance $d\left(r\right)$ where $k=\xi/d\left(r\right)$; the resulting improved Reissner-Nordstr\"{o}m geometry reads as follows\footnote{For more details see section 2}:
\begin{eqnarray}
ds^{2}=-f^{I}(r)dt^{2}+f^{I}(r)^{-1}dr^{2}+r^{2}d\Omega^{2}, 
\nonumber \\
f^{I}(r) \equiv 1-\frac{2G(r)M}{r}+\frac{G(r) Q^{2}}{r^{2}}, \label{I_RN}
\end{eqnarray}
\noindent where the super-index $I$ stands for ``improved'' and the running Newton constant is given by\footnote{Expression (\ref{Gk}) is the result of solving the \textit{exact} renormalization group flow equation for quantum Einstein gravity derived by M. Reuter in 1996 \cite{Reuter_96}, in the frame of the so-called Einstein-Hilbert truncation. In this sense, expression (\ref{Gk}) is a non-perturbative result for the running Newton constant. In section 2 we describe the main steps of the derivation of (\ref{Gk}); for further details see \cite{Reuter_96,Bonanno-Reuter}. }   
\begin{eqnarray}
G(k)&\equiv& \frac{G_{0}}{1+wG_{0}k^2}, \label{Gk} \\
G(r)&\equiv& G\left(k=\xi/d\left(r\right)\right)= \frac{G_{0}d^{2}(r)}{d^{2}(r)+\bar{w}G_{0}},  \label{GC}
\end{eqnarray}
\noindent here $\bar{w}\equiv w\xi^2 $ is a free parameter that turns ``on'' and ``off'' the quantum effects. 
The applicability of (\ref{GC}) and (\ref{I_RN}) corresponds, as already mentioned, to sufficiently long distances related to the  Planck length $l_{p}=\sqrt{\frac{\hbar G_{0}}{c^{3}}}$.

This article has the following structure: in section $2$ we summarize the theoretical basis and main steps in the framework of the Einstein-Hilbert truncation that lead to a renormalization group improvement of classical solutions of Einstein field equations (for more detail see \cite{Reuter_96,Saueressig,Bonanno-Reuter}). In section $3$ we present the main results of our work \cite{Ruiz}: we study the critical surfaces and extremality conditions of the renormalization group improved Reissner-Nordstr\"{o}m spacetime (\ref{I_RN}) that describes a charged non-rotating black hole. We also address the structural stability of these surfaces as roots of polynomials with the variation of the quantum parameter $\bar{w}$ \cite{Tuiran,Catastrofe}. We derive an expression for the area of the event horizon of the improved metric; we compute the surface gravity at the horizon, via the proportionality constant between the Killing vector for the time symmetry and the derivative of its squared norm \cite{Poisson} and we show that an exact differential can be built with the surface gravity as integrating factor; we also calculate the Komar integral for the mass of the black hole at the event horizon and we compare it with the asymptotic value at infinity, looking for signatures of anti-screening due to the gravitational quantum fluctuations for the long-range regime. In section 4 we present our conclusions, perspectives and comparison to related works.
\section{Renormalization Group Improvement of Classical Spacetimes}
\label{S:2}
\subsection{The Running Newton Constant}
\label{sec:command}
The effective average action $\Gamma_{k}$ is the result of integrating out higher momentum modes down to an infrared cutoff, in the generating functional for Quantum Einstein Gravity (QEG), which is defined as an euclidean path integral in the metric field for a classical fundamental action $S$. The process of integrating out modes, is carried out through the inclusion of a cutoff term $R_{k}$ that supresses all modes $p<k$ in the path integral. This process can be understood as a specific continuum implementation of the Wilsonian renormalization group transformations. In this way $\Gamma_{k}$ interpolates between the effective action $\Gamma$ for $k\rightarrow 0$ and $S$ the ``bare'' action for $k\rightarrow \infty$. 
\noindent \\ The infinitesimal change of $\Gamma_{k}$ with $k$ is described by a functional differential equation, the so-called Exact Renormalization Group Equation (ERGE)
\begin{equation}
k\partial_{k}\Gamma_{k}=\frac{1}{2}Tr\left[k\partial_{k}R_{k}\left(\Gamma_{k}^{\left(2\right)}+R_{k}\right)^{-1}\right],  
\label{EGR}
\end{equation}
\noindent where $\Gamma^{\left(2\right)}_k$ is the hessian matrix of functional derivatives of $\Gamma_k$ with respect to the dynamical fields. For a detailed discussion of this equation see \cite{Reuter_96}. In general it is impossible to find exact solutions to (\ref{EGR}). A useful method to find approximate non-perturbative solutions, consists in carrying out truncations in the theory space; where the exact RG-flow is projected onto a finite dimensional subspace. In practice, an ansatz for $\Gamma_{k}$ with a finite set of invariants is inserted in (\ref{EGR}). This leads to a finite set of differential equations of the form:
\begin{equation}
k\partial_{k}g_{i}(k)=\beta_{i}(g_{1},g_{2},\cdots),
\label{FBet}
\end{equation}
\noindent where $g_{i}(k)$ are dimensionless couplings for each invariant in the truncation.   
\noindent In this work we concentrate in the Einstein-Hilbert truncation \cite{Reuter_96, Saueressig} that restricts the study of the evolution of $\Gamma_{k}$ to the two-dimensional subspace expanded by the Newton constant $G_{k}$ and the cosmological constant $\bar{\lambda}_{k}$. The ansatz for $\Gamma_{k}$ in this truncation is given by\footnote{For a detailed study of the more general scheme of QEG with running Yang-Mills couplings see \cite{DHR1, DHR2,Harst,Eich-Verst}. We briefly analyse in our discussion of section 4 some recent results for the improved Reissner-Nordstr\"om metric under this approach.}
\begin{equation}
\Gamma_{k}\left[g_{\mu\nu}\right]=\left(16\pi G_{k}\right)^{-1}\int
d^{d}x\sqrt{g}\left\{-R\left(g\right)+2\bar{\lambda}_{k}\right\},
\label{T E-H}
\end{equation}
\noindent where we have generalized to $d$ dimensions. After substituting (\ref{T E-H}) in the ERGE equation (\ref{EGR}) one obtains a system of coupled differential equations for the dimensionless Newton and cosmological constants $g_{k}\equiv k^{d-2}G_{k}$ and $\lambda _{k}\equiv k^{-2}\bar{\lambda}_{k}$ respectively, given by \cite{Reuter_96, Saueressig, Bonanno-Reuter}:
\begin{equation}
\partial_{t}g=\left(d-2+\eta _{N}\right)g , 
\label{Eg}
\end{equation}
\noindent and 
\begin{eqnarray}
\partial_{t}\lambda=&-\left( 2-\eta_{N}\right)\lambda+\frac{1}{2}g\left(4\pi\right)^{\left(1-\frac{d}{2}\right)}\times \nonumber \\
&\times\left[2d\left(d+1\right)\Phi_{\frac{d}{2}}^{1}\left(-2\lambda\right)-8d\Phi_{\frac{d}{2}}^{1}\left(0\right)-d\left(d+1\right)\eta_{N}\tilde{\Phi}_{\frac{d}{2}}^{1}\left(-2\lambda\right)\right],\label{EL}
\end{eqnarray}
\noindent here $t\equiv\ln k$ is the evolution parameter, the so called renormalization time. The function $\eta_{N}\left(g,\lambda\right)$ is defined by
\begin{equation}
\eta_{N}\left(g,\lambda\right)=\frac{gB_{1}\left(\lambda\right)}{1-gB_{2}\left(\lambda\right)}\;,  
\label{DA}
\end{equation}
 \noindent and functions $B_{1}\left(\lambda \right)$ and $B_{2}\left(\lambda\right)$ are given by
\vspace{-0.1cm}
\begin{eqnarray} 
B_{1}\left( \lambda\right)\equiv&\frac{1}{3}\left(4\pi \right)^{\left(1-\frac{d}{2}\right)}\times  \nonumber\\ 
&\times\left[d\left(d+1\right)\Phi_{\frac{d}{2}-1}^{1}\left(-2\lambda\right)-6d\left(d-1\right)\Phi_{\frac{d}{2}}^{2}\left(-2\lambda\right)-4d\Phi_{\frac{d}{2}-1}^{1}\left(0\right)-24\Phi_{\frac{d}{2}}^{2}\left(0\right)\right],  \nonumber\\
B_{2}\left(\lambda\right)\equiv&-\frac{1}{6}\left(4\pi \right)^{\left(1-\frac{d}{2}\right)}\left[d\left(d+1\right)\tilde{\Phi}_{\frac{d}{2}-1}^{1}\left(-2\lambda \right)-6d\left(d-1\right)\tilde{\Phi}_{\frac{d}{2}}^{2}\left(-2\lambda\right)\right],
\label{B1B2}
\end{eqnarray}
\noindent where the threshold functions for $p=1,2,...$ given by
\begin{eqnarray}
\Phi_{n}^{p}\left(s\right)&=\frac{1}{\Gamma\left(n\right)}\int_{0}^{\infty}dz\;z^{n-1}\frac{R^{\left(0\right)}\left(z\right)-zR^{\left(0\right)\prime}\left(z\right)}{\left[z+R^{\left(0\right)}\left(z\right)+s\right]^{p}}\;, \label{umbral0}\\
\tilde{\Phi}_{n}^{p}\left(s\right)&=\frac{1}{\Gamma \left( n\right)}\int_{0}^{\infty }dz\;z^{n-1}\frac{R^{\left(0\right)}\left(z\right)}{\left[z+R^{\left(0\right)}\left(z\right)+s\right]^{p}}\;, 
\label{umbral} 
\end{eqnarray}
\noindent depend on the cutoff function $R^{\left(0\right)}\left(z\right)$ with 
$z\equiv p^{2}/k^{2}$. $R^0$ is arbitrary, except for the two conditions $R^{\left(0\right)}\left(0\right) =1$ and $R^{\left(0\right)}\left(z\right)\rightarrow 0$ for $z\rightarrow\infty$. For explicit calculations we choose the exponential form:
\begin{equation}
R^{\left(0\right)}\left(z\right)=\frac{z}{e^{z}-1}\;.  
\label{FC}
\end{equation}
\noindent From now on we assume that $\bar{\lambda}$ $\ll k^{2}$ for our scales of interest, this means that $\lambda\left(k\right)\approx 0$ and we do not consider the influence of the running cosmological constant in the physics of spherically symmetric black holes \footnote{For a detailed study of this subject see \cite{Saueressig-1}}. As a consequence the evolution of $g$ is entirely governed by
\begin{equation}
k\partial_{k}g=\left(2+\eta_{N}\right)g=\beta\left(g\left(k\right)\right),  
\label{Eg1}
\end{equation}
\noindent where the function $\eta_{N}\left(g\right)$ is given by
\begin{equation}
\eta_{N}\left(g\right)=\frac{gB_{1}}{1-gB_{2}}\;,  
\label{DA1}
\end{equation}
\noindent with 
\begin{equation}
B_{1}\equiv B_{1}\left(0\right)=-\frac{1}{3\pi}\left[24\Phi_{2}^{2}\left(0\right)-\Phi_{1}^{1}\left(0\right)\right], 
\label{Def B1}
\end{equation}
and
\begin{equation}
B_{2}\equiv B_{2}\left(0\right)=\frac{1}{6\pi}\left[18\tilde{\Phi}_{2}^{2}\left(0\right) -5\tilde{\Phi}_{1}^{1}\left(0\right)\right]. 
\label{Def B2}
\end{equation}
\noindent Substituting the cutoff function (\ref{FC}) in definitions (\ref{umbral0}) and (\ref{umbral}) leads to the results
\begin{eqnarray*}
\Phi_{1}^{1}\left(0\right)&=\frac{\pi^{2}}{6},\qquad \Phi_{2}^{2}\left(0\right)=1 \\
\tilde{\Phi}_{1}^{1}\left(0\right)&=1,\qquad \; \; \, \tilde{\Phi}_{2}^{2}\left(0\right)=\frac{1}{2}
\end{eqnarray*}
\noindent and, 
\begin{equation}
B_{1}=\frac{\pi}{18}-\frac{8}{\pi},\; B_{2}=\frac{2}{3\pi}\,,
\end{equation}
\noindent with these expressions for $B_1$ and $B_2$, we substitute (\ref{DA1}) in (\ref{Eg1}) to obtain the following expression for the $\beta$-function
\begin{equation}
\beta\left(g\right)=2g\left(\frac{1-\left(B_{2}-\frac{1}{2}B_{1}\right)g}{1-B_{2}g}\right),  
\label{Eg2}
\end{equation}
\noindent with the following definitions
\begin{equation}
w\equiv-\frac{1}{2}B_{1},\; \omega^{\prime}=w+B_{2}\,, 
\label{Def w Omega}
\end{equation}
\noindent as a result, the $\beta$-function takes the form 
\begin{equation}
\beta\left(g\right)=2g\left(\frac{1-\omega^{\prime}g}{1-B_{2}g}\right),  
\label{Eg2_r}
\end{equation}
\noindent where
\begin{equation}
w=\frac{4}{\pi}\left(1-\frac{\pi^{2}}{144}\right) ,\;\omega^{\prime}=\frac{14}{3\pi}-\frac{\pi}{36}\;.
\label{Eg3}
\end{equation}
\noindent The evolution equation (\ref{Eg1}) for $g(k)$ with $\beta$ in (\ref{Eg2_r}) leads to the existence of two fixed points (let's call them $g_{\ast}$) as we can check from the condition $\beta\left(g_{\ast }\right)=0$. One of the solutions is an IR-attractive gaussian fixed point $g_{\ast}^{\text{IR}}=0$, the other one is a UV-attractive non-gaussian fixed point: 
\begin{equation*}
g_{\ast}^{\text{UV}}=\frac{1}{\omega^{\prime}}\;. \\
\end{equation*}
Figure 1 shows the dependence on $g$ of the $\beta$-function. The ultraviolet fixed point separates a weak coupling regime for $(g<g_{\ast}^{\text{UV}})$ from a strong coupling regime for $(g>g_{\ast}^{\text{UV}})$. 
\begin{figure}[H]
\centering
\includegraphics[scale=0.8]{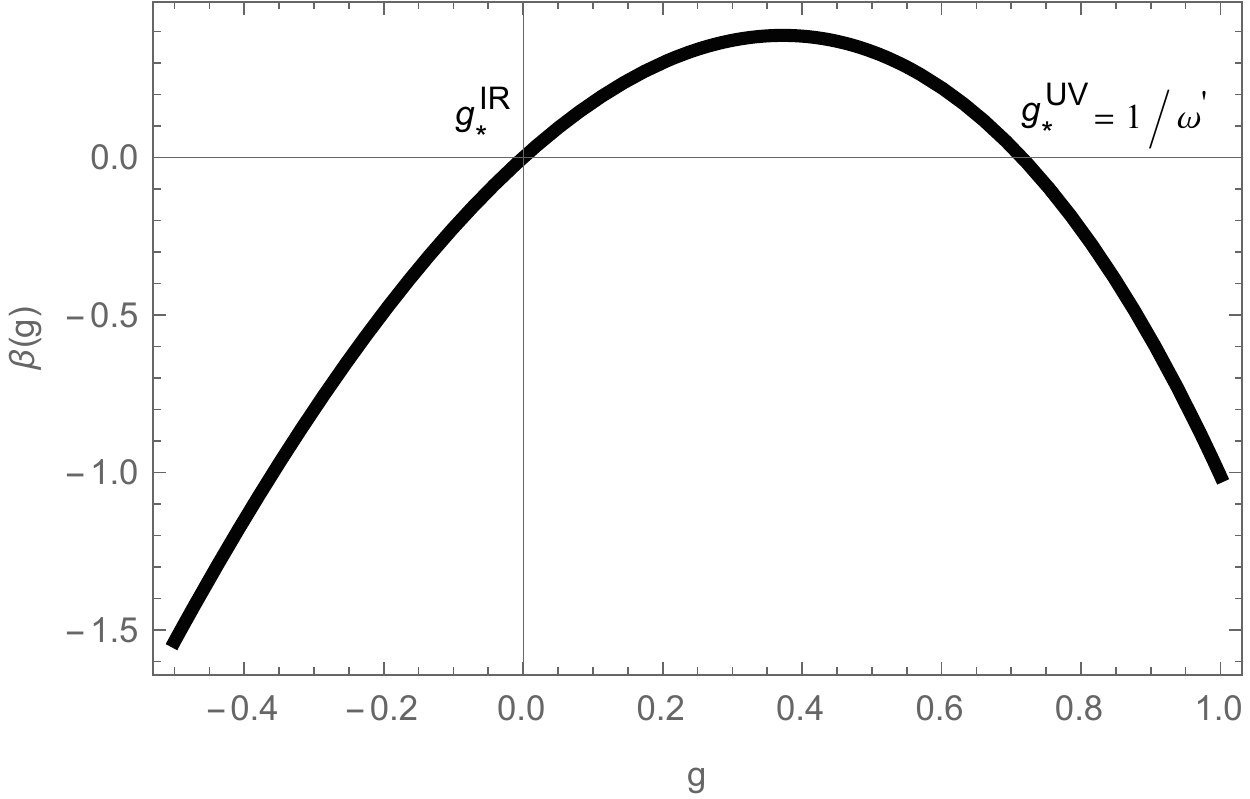}
\caption{\small{ The function $\beta(g)$ for the dimensionless Newton constant.}}
\label{fig:Fig. 1}
\end{figure}
\noindent Since the $\beta$-function (\ref{Eg2}) is positive for $g\in\left[0, g_{\ast}^{\text{UV}}\right]$ and negative in other case, the RG trajectories $g(k)$ for the dimensionless Newton constant fall into three classes\cite{Bonanno-Reuter,Souma}:
\begin{enumerate}
\item Trajectories where $g\left(k\right)<0$ for all $k$, flowing to $g_{\ast}^{\text{IR}}=0$ for $k\rightarrow 0$.
\item Trajectories where $g\left(k\right)>g_{\ast }^{\text{UV}}$ for all $k$, flowing to $g_{\ast}^{\text{UV}}=1/\omega^{\prime}$ for $k\rightarrow\infty$.
\item Trajectories where $g\left(k\right)\in\left[0, g_{\ast}^{\text{UV}}\right]$ for all $k$, flowing to $g_{\ast }^{\text{IR}}\;$ for $\;k\rightarrow 0$ and $g_{\ast}^{\text{UV}}$ for $k\rightarrow\infty$.
\end{enumerate}
\noindent The first class imply a negative Newton constant, and the second class is isolated from a low energy regime; as a result, only the third class of trajectories turns out to be relevant for the present work.
\noindent The differential equation (\ref{Eg1}) with the $\beta$-function (\ref{Eg2}) can be integrated analytically; the result is given by \cite{Bonanno-Reuter} 
\begin{equation}
\frac{g}{\left(1-\omega^{\prime}g\right)^{\frac{w}{\omega^{\prime}}}}=\frac{g\left(k_{0}\right)}{\left[1-\omega^{\prime}g\left(k_{0}\right)\right]^{\frac{w}{\omega^{\prime}}}}\left(\frac{k}{k_{0}}\right)^{2}.
\label{g}
\end{equation}
This expression cannot be solved in closed form for $g=g\left(k_{0}\right)$. In order to find an approximate analytical expression for the running of the Newton constant, we notice that the fraction $\frac{\omega^{\prime}}{w}$ computed from (\ref{Eg3}) is near to unity $\left(\frac{\omega^{\prime}}{w}\approx 1.18\right)$. Substituting $\frac{\omega^{\prime}}{w} = 1$ in equation (\ref{g}) leads to a good approximation with all qualitative features expected for the running $g\left(k\right)$ \cite{Bonanno-Reuter}:
\begin{equation}
g\left(k\right)=\frac{g\left(k_{0}\right)k^{2}}{wg\left(k_{0}\right)k^{2}+\left[1-wg\left(k_{0}\right)\right]k_{0}^{2}}\;.
\label{g1}
\end{equation}
\noindent The fully dimensional Newton's constant $G\left(k\right)\equiv g\left(k\right)/k^{2}$ turns out to be
\begin{equation}
G\left(k\right)=\frac{G\left(k_{0}\right)}{1+w G\left(k_{0}\right)\left[k^{2}-k_{0}^{2}\right]}\;. 
\label{G}
\end{equation}
\noindent We set $k_{0}=0$ as a scale of reference for which one identifies $G_{0}\equiv G\left(k_{0}=0\right)$ with the current measured value of the Newton constant. As a result we have
\begin{equation}
G\left(k\right)=\frac{G_{0}}{1+wG_{0}k^{2}}\;,  
\label{G1}
\end{equation}
\noindent where we leave $w$ without evaluation, since it is a constant that depends explicitly on the specific choice of the cutoff function $R^{(0)}$. We say in this sense that $w$ is a non-universal quantity. In order to express $G(k)$ as an observable quantity, a second source of non-universality must appear, so that it compensates the dependence on $R^{(0)}$ of $w$. This issue will be analyzed in the next section.
\noindent As mentioned before, the approximation $\frac{\omega^{\prime}}{w} \approx 1$ still preserves the expected qualitative features of the running Newton constant. From (\ref{G1}) we see for example that for $k \to 0$ we have 
\begin{equation}
G\left(k\right)=G_{0}-wG_{0}^{2}k^{2}+O\left(k^{4}\right),
\end{equation}
\noindent namely, we recover $G_{0}$; whereas for $k^{2}>>G_{0}^{-1}$ we have $G\left(k\right)\approx \frac{1}{wk^{2}}$; this means that for $k\to\infty$, Newton constant vanishes; this is consistent with the result for the running $G$ near to the Planck scale, in other scenarios (see \cite{Polyakov})
\subsection{Cutoff Identification}
\label{sec:math}
Following the spirit of the derivation with renormalization group of the Uehling correction to the Coulomb potential in QED, where the scale $k$ is identified with the inverse of the radial distance $r$ \cite{Ditt-Reuter,Uehling}, we introduce an analogue cutoff identification that relates $k$ with the geometry of space-time \cite{Bonanno-Reuter,Reuter_1}:
\begin{equation}
k(P)=\frac{\xi}{d(P)}\;,
\label{Cutoff}
\end{equation} 
\noindent where $\xi$ is a constant that represents our lack of knowledge about the exact physical mechanism leading to the infrared cutoff, and $d(P)$ is an invariant radial distance replacing the coordinate dependent parameter $r$; as required by general relativity. It can be defined as the proper distance between an initial point, fixed in space-time $P_{0}$ (usually the origin of coordinates) and a simultaneous final point $P$. The proper distance $d(P)$ is the result of integrating the infinitesimal proper length $\sqrt{|ds^{2}|}$ along a definite path $\mathcal{C}$ which is usually chosen to be a straight line in 3-D space \cite{Bonanno-Reuter}:
\begin{equation}
d(P)=\int_{\mathcal{C}}\sqrt{|ds^{2}|}\;.
\label{d(P)}
\end{equation}
\noindent Now substituting (\ref{Cutoff}) in (\ref{G1}) leads to
\begin{equation}
G(P)=\frac{G_{0}d(P)^{2}}{d(P)^{2}+G_{0}\bar{w}}\; , \; \bar{w} \equiv w \xi^2 \label{G2}
\end{equation}
\noindent where $w$ and $\xi$ appear together in the form of the product $\bar{w} \equiv w\xi^{2}$. The new parameter $\bar{w}$ cannot be obtained exclusively from renormalization group arguments. In principle, it should be experimentally determined, after a measurement of the quantum correction to the Newton potential \cite{Donoghue}. It fulfills the following properties that defines it as a good parameter to switch on and off the quantum effects:
\begin{enumerate}
\item $\bar{w}$ is proportional to $\hbar$.
\item $\bar{w}$ is the only constant in (\ref{G2}) related to the evolution of $G$ with the scale $k$.
\item When $\bar{w}=0$ we recover the usual value of the Newton constant with $G(P)=G_{0}$.
\end{enumerate}
Even though expression (\ref{d(P)}) depends in general on the election of the path of integration $\mathcal{C}$; if we choose a straight path between the origin and $P$ it is straightforward to conclude that for a spherically symmetric space-time, the proper distance $d(P)$ depends exclusively on the radial coordinate $r$; since the integrand in (\ref{d(P)}) depends through the components of the metric, only on $r$ (for more details see \cite{Bonanno-Reuter,Tuiran}). As a result we set $d(P)=d(r)$ for spherically symmetric space-times, and the expression for the cutoff identification takes the form\\
\begin{equation}
k=\frac{\xi}{d(r)}\;. \label{Cutoff1}
\end{equation}
\noindent Substituting (\ref{Cutoff1}) in expression (\ref{G1}) for $G(k)$ leads to the following formula for $G(r)$:
\begin{equation}
G(r)=\frac{G_{0}d(r)^{2}}{d(r)^{2}+\bar{w}G_{0}}. \label{G3}
\end{equation}
For the classical Reissner-Nordstr\"{o}m metric (\ref{Class RN}) with $\left(t,\theta,\phi\right)=\text{const} $, $d(r)$ takes the following ``piecewise" form after integrating (\ref{d(P)}):
\begin{equation}
d(r)=\left\{
\begin{array}{ll}
d_{1}(r)\quad \rm if\ r<r_{-}\\ 
d_{2}(r)\quad \rm if\ r_{-}<r<r_{+}\\ 
d_{3}(r)\quad \rm if\ r_{+}<r\:,
\end{array}
\right. \\ \label{d_r_1}
\end{equation}

\begin{equation}
r_{\pm}=m\pm\sqrt{m^{2}-Q^{2}}\:,
\end{equation}

\begin{equation}
d_{1}(r)=\sqrt{r^{2}+Q^{2}-2mr}+m\:\rm ln\left(\frac{-r+m-\sqrt{r^{2}+Q^{2}-2mr}}{|Q-m|}\right)-Q\:,
\end{equation}

\begin{eqnarray}
\begin{aligned}
d_{2}(r)=&\frac{m}{2}\:\rm ln\bigg|\frac{m+Q}{m-Q}\bigg|-Q-\sqrt{2mr-r^{2}-Q^{2}}\\
&+m\:arctan\left(\frac{r-m}{\sqrt{2mr-r^{2}-Q^{2}}}\right)+\frac{m\pi}{2}\:,
\end{aligned}
\end{eqnarray}

\begin{eqnarray}
\begin{aligned}
d_{3}(r)=&\sqrt{r^{2}+Q^{2}-2mr}+m\:\rm ln\left(r-m+\sqrt{r^{2}+Q^{2}-2mr}\right)\\
&+m\pi-Q-m\:\rm ln|m-Q|\:.
\end{aligned}
\label{d3}
\end{eqnarray}
Figure 2 shows the radial function $d(r)$ in (\ref{d_r_1}) for several values of the charge $Q$ of the Reissner-Nordstr\"{o}m black hole. The $Q=0$ case coincides with $d(r)$ for the Schwarzschild spacetime. The asymptotic behavior $d(r)\to r +m\ln{r} + \text{const} $ for $r\to \infty$ directly implied from expression (\ref{d3}) can be observed by comparing these curves with the $d(r)=r$ dashed straight line. For large $r$, the line $d(r)=r$  turns out to be the dominant asymptotic behavior.       
\begin{figure}[H]
\centering
\includegraphics[scale=0.52]{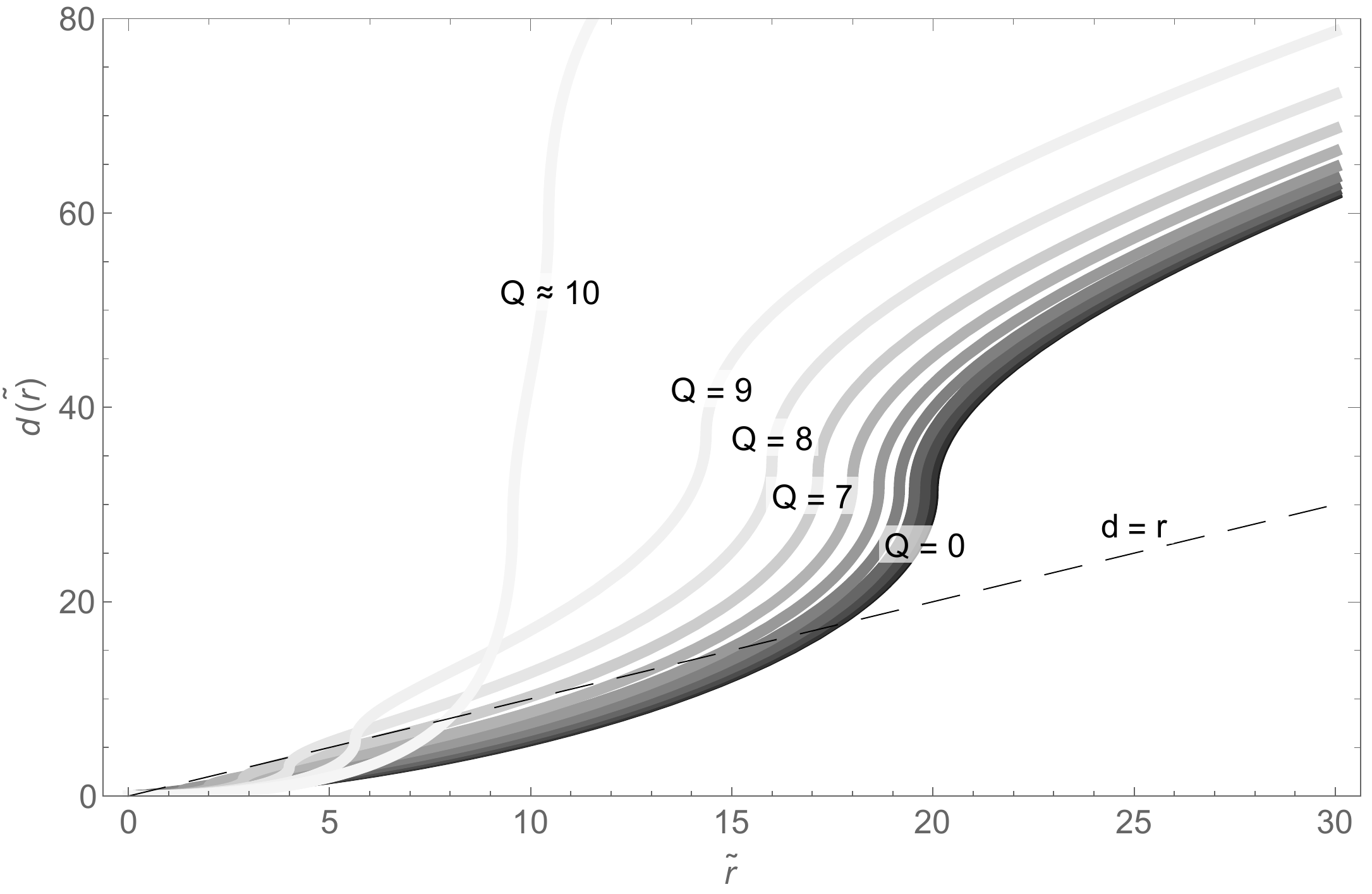}
\caption{\small{ The radial function $d(r)$ for the Reissner-Nordstr\"{o}m spacetime.}}
\label{fig:Fig. 2}
\end{figure}

\noindent If the phenomenon to be analysed takes place far enough from the singularity $(r>>0)$; namely the region where our present method of improving classical space-times is understood to be mostly reliable; then the above mentioned asymptotic behavior of $d(r)$ approaching $r$ can be applied \cite{Bonanno-Reuter, Tuiran}. As a consequence we approximate:
\begin{equation}
d(r)=r,\; k\approx\frac{\xi}{r}\;,\; r>>0,
\label{Cutoff2}
\end{equation}
\noindent and replacing $d(r)=r$ in (\ref{G3}) leads to the following expression for $G(r)$:
\begin{equation}
G(r)=\frac{G_{0}r^{2}}{r^{2}+\bar{w}G_{0}}\;. 
\label{G4}
\end{equation}
\noindent Substituting (\ref{G4}) in the expression (\ref{I_RN}) for $f^{I}(r)$ leads to
\begin{equation}
f^{I}(r)=1-\frac{2G_{0}rM}{r^{2}+\bar{w}G_{0}}+\frac{G_{0}Q^{2}}{r^{2}+\bar{w}G_{0}}\;,
\label{F_I_r_large}
\end{equation}
which turns out to define, after substituted in the squared length (\ref{I_RN}), the improved version of Reissner-Nordstr\"{o}m space-time for large $r$. Expressions (\ref{G4}) and (\ref{F_I_r_large}) will be applied in sections (3.1) and (3.3) where we study the leading effects of the improvement in the event horizon surfaces and the mass inside the black hole. 
The rest of this work (section (3.2)) is concerned about the effects of the improvement on the first law of black hole dynamics; and in this case we will simply use an undefined function $G(r)$, in order to derive generic features independent of the $r \to \infty$ asymptotic behavior of the Newton constant. As a result, we will stay with expression (\ref{I_RN}) for $f^{I}(r)$:
\begin{equation}
f^{I}(r)=1-\frac{2G(r)M}{r}+\frac{G(r)Q^{2}}{r^{2}} \label{FIC}.
\end{equation}
\noindent With expressions (\ref{F_I_r_large}) or (\ref{FIC}) we have the starting point to analyze the different properties of the improved Reissner-Nordstr\"{o}m spacetime. 
\section{Results}
\subsection{Critical Surfaces for the $d(r)=r$ Approximation}

\noindent The condition $f^{I}(r)=0$ for event horizons leads, in the most general way to the following equation:
\begin{equation}
r^{2}-2G(r)Mr+G(r)Q^{2}=0. \label{PC_gral}
\end{equation}
In the $d(r)=r$ approximation, equation (\ref{PC_gral}) turns out to be a quadratic polynomial given by 
\begin{equation}
r^{2}-2G_{0}Mr+\bar{w}G_{0}+G_{0}Q^{2}=0, \label{PC}
\end{equation}
\noindent where we have used expression (\ref{G4}). The solutions of (\ref{PC}) are given by%
\begin{equation}
r_{\pm }^{I}=G_{0}\left( M\pm \sqrt{M^{2}-\frac{\bar{w}}{G_{0}}-\frac{Q^{2}}{G_{0}}}\right).
\end{equation}
\noindent By applying the definitions given in appendix A, a dimensionless form of the horizon equation (\ref{PC}) can be derived:
\begin{equation}
P^{\bar{w}}_{\widetilde{Q}}(\widetilde{r})\equiv \widetilde{r}^{2}-2\widetilde{m}\widetilde{r}+\bar{w}+\widetilde{Q}^{2}=0. \label{PCA}
\end{equation}
\noindent From equation (\ref{PCA}) one can identify the following cases:
\begin{enumerate}
	\item Critical surfaces of classic Schwarzschild $(\bar{w}=0, \widetilde{Q}=0)$.
	\item Critical surfaces of classic Reissner-Nordstr\"{o}m $(\bar{w}=0, \widetilde{Q}\neq0)$.
	\item Critical surfaces of improved Schwarzschild $(\bar{w}\neq0, \widetilde{Q}=0)$.
	\item Critical surfaces of improved Reissner-Nordstr\"{o}m $(\bar{w}\neq0, \widetilde{Q}\neq0)$.
\end{enumerate}
\subsection*{Case 1}
\noindent The case that results from the configuration $\bar{w}=0$ and $\widetilde{Q}=0$ leads to the well-known Schwarzschild's singularity . The equation (\ref{PCA}) is reduced to:
\begin{equation}
P^{0}_{0}(\widetilde{r})\equiv\widetilde{r}(\widetilde{r}-2\widetilde{m})=0. \label{S}
\end{equation}
\noindent Clearly the solutions of (\ref{S}) are $\widetilde{r}=0$, the singularity at the origin, or $\widetilde{r}=2\widetilde{m}$ the coordinate singularity that defines the event horizon.

\subsection*{Case 2}
\noindent The case defined by $\bar{w}=0, \widetilde{Q}\neq0$ turns equation (\ref{PCA}) into the following one:
\begin{equation}
P^{0}_{\widetilde{Q}}(\widetilde{r})\equiv\widetilde{r}^{2}-2\widetilde{m}\widetilde{r}+\widetilde{Q}^{2}=0. \label{R}
\end{equation}

\noindent The solutions of (\ref{R}) are 
\begin{equation}
\widetilde{r}_{\pm}=\widetilde{m}\pm\sqrt{\widetilde{m}^{2}-\widetilde{Q}^{2}}.
\label{SR}
\end{equation}
\noindent The dimensional form of these solutions can be found by multiplying $\widetilde{r}$ by $l_{pl}=\sqrt{G_{0}}$, namely:
\begin{equation}
r_{\pm}=MG_{0}\pm\sqrt{(MG_{0})^{2}-G_{0}Q^{2}}.
\end{equation}

\noindent Or substituting  $m=MG_{0}$ leads to:
\begin{equation}
r_{\pm}=m\pm\sqrt{m^{2}-G_{0}Q^{2}},
\end{equation}

\noindent which corresponds to the usual form of the event horizon solutions of the classical Reissner-Nordstr\"{o}m spacetime.

\subsection*{Case 3}
The case with $\bar{w}\neq0, \widetilde{Q}=0$ gives the critical surfaces of the improved  Schwarzschild spacetime. In this case the equation (\ref{PCA}) is reduced to
\begin{equation}
P^{\bar{w}}_{0}(\widetilde{r})\equiv\widetilde{r}^{2}-2\widetilde{m}\widetilde{r}+\bar{w}=0, \label{SC}
\end{equation}

\noindent which is equivalent to (\ref{R}) if the identification $\widetilde{Q}^{2} \rightarrow \bar{w}$ is done. As a result, the solutions of (\ref{SC}) have a similar form to the solutions of (\ref{R}). In this case we have again the two event horizons obtained by Bonanno and Reuter in \cite{Bonanno-Reuter} for the improved Schwarzschild spacetime; they are given by:
\begin{equation}
\widetilde{r}^{I}_{Sch\pm}=\widetilde{m}\pm\sqrt{\widetilde{m}^{2}-\bar{w}}. \label{RSC}
\end{equation}

\noindent Multiplying (\ref{RSC}) by $l_{pl}=\sqrt{G_{0}}$ leads to a  dimensional form for the improved Schwarzschild event horizons
\begin{equation}
r^{I}_{Sch\pm}=MG_{0}\pm\sqrt{(MG_{0})^{2}-G_{0}\bar{w}},
\end{equation}

\noindent or equivalently with $m=MG_{0}$:
\begin{equation}
r^{I}_{Sch\pm}=m\pm\sqrt{m^{2}-G_{0}\bar{w}}.
\end{equation}

\noindent The existence of two different radii $r^{I}_{Sch\pm}$ represents the splitting due to the correction with renormalization group, of the unique Schwarzschild event horizon, in a set of two horizons that has already been discussed in \cite{Bonanno-Reuter}.

\subsection*{Case 4}
The more general case $\bar{w}\neq0, \widetilde{Q}\neq0$ is associated to the improved Reissner-Nordstr\"{o}m spacetime. From the quadratic equation in (\ref{PCA}) two solutions are found:
\begin{equation}
\widetilde{r}_{\pm}^{I}=\widetilde{m}\pm\sqrt{\widetilde{m}^{2}-\bar{w}-\widetilde{Q}^{2}}. \label{RRNC}
\end{equation}
\noindent In the case $\widetilde{m}^{2}>\bar{w}+\widetilde{Q}^{2}$ there exist two horizons as in the previous cases; they coalesce in just one solution when $\widetilde{m}^{2}=\bar{w}+\widetilde{Q}^{2}$, the so called extremal configuration, and further disappear if $\widetilde{m}^{2}<\bar{w}+\widetilde{Q}^{2}$.

\noindent Similarly to the previous cases, a dimensional expression is found after multiplying (\ref{RRNC}) with $l_{pl}=\sqrt{G_{0}}$:
\begin{equation}
r_{\pm}^{I}=MG_{0}\pm\sqrt{(MG_{0})^{2}-G_{0}\bar{w}-G_{0}Q^{2}} \label{Imp_EH_RN_0},
\end{equation}

\noindent or in terms of the geometric mass $m\equiv MG_{0}$ one gets:
\begin{equation}
r_{\pm}^{I}=m\pm\sqrt{m^{2}-G_{0}\bar{w}-G_{0}Q^{2}}\label{Imp_EH_RN} .
\end{equation}

\noindent This is an important result obtained in this work. It corresponds to the event horizon solutions for the improved Reissner-Nordstr\"{o}m spacetime. In the following sections we will address the issues of stability and extremality of these solutions.
\subsubsection{Condition of quantum extremality for the improved Reissner-Nordstr\"{o}m horizon}
The condition of extremality for the classical Reissner-Nordstr\"{o}m spacetime is given when the two solutions $\widetilde{r}_{\pm}$ are degenerated to a single, $\widetilde{r}_{+}=\widetilde{r}_{-}$. This happens when $\widetilde{m}=\widetilde{Q}$ and as a result $\widetilde{r}_{ext}=\widetilde{m}=\widetilde{Q}$, as can be seen from (\ref{SR}). Similarly the condition of extremality can be generalized for improved Reissner-Nordstr\"{o}m spacetime, namely $\widetilde{r}_{+}^{I}=\widetilde{r}_{-}^{I}$. This occurs when $\widetilde{m}=\sqrt{\bar{w}+\widetilde{Q}^{2}}$ and as a result $\widetilde{r}_{ext}^{I}=\widetilde{m}=\sqrt{\bar{w}+\widetilde{Q}^{2}}$, as can be seen from (\ref{RRNC}).

Figure $3$ shows the dependence $\widetilde{m}(\widetilde{Q})$ given by the condition of quantum extremality for improved Reissner-Nordstr\"{o}m extreme black hole with $d(r)=r$. The dashed line represents the dependence $\widetilde{m}(\widetilde{Q})=\widetilde{Q}$ of classical Reissner-Nordstr\"{o}m spacetime. A new type of extremal configuration related to the parameter $\bar{w}$ appears as a result from the improvement. This configuration is reached at the Planck scale where $\widetilde{m} = \sqrt{\bar{w}}$, and it has to be distinguished from the state reached at $\widetilde{m} = \widetilde{Q}$ for the extremal classical Reissner-Nordstr\"{o}m black hole. In contrast, the similarity to the extremal configuration presented in \cite{Tuiran} for the mass $\widetilde{m}(\widetilde{a})$ as a function of angular momentum of the improved Kerr black hole is clear. At the Planck scale, for $\widetilde{Q} \rightarrow 0$, $\widetilde{m}$ assumes its minimum value at $\sqrt{\bar{w}}$, and for $\widetilde{Q} \rightarrow \infty , \widetilde{m}$ approaches the classical extremal behavior, namely $\widetilde{m} \rightarrow \widetilde{Q}$. The deviation from the classical behavior when $\widetilde{Q}$ approaches the Planck scale can be interpreted as a possible prediction of the existence of an ``inert'' final state, in the process of evaporation, where the mass of the black hole does not completely disappear \cite{Bonanno-Reuter}.
\begin{figure}[H]
\centering
\includegraphics[scale=0.8]{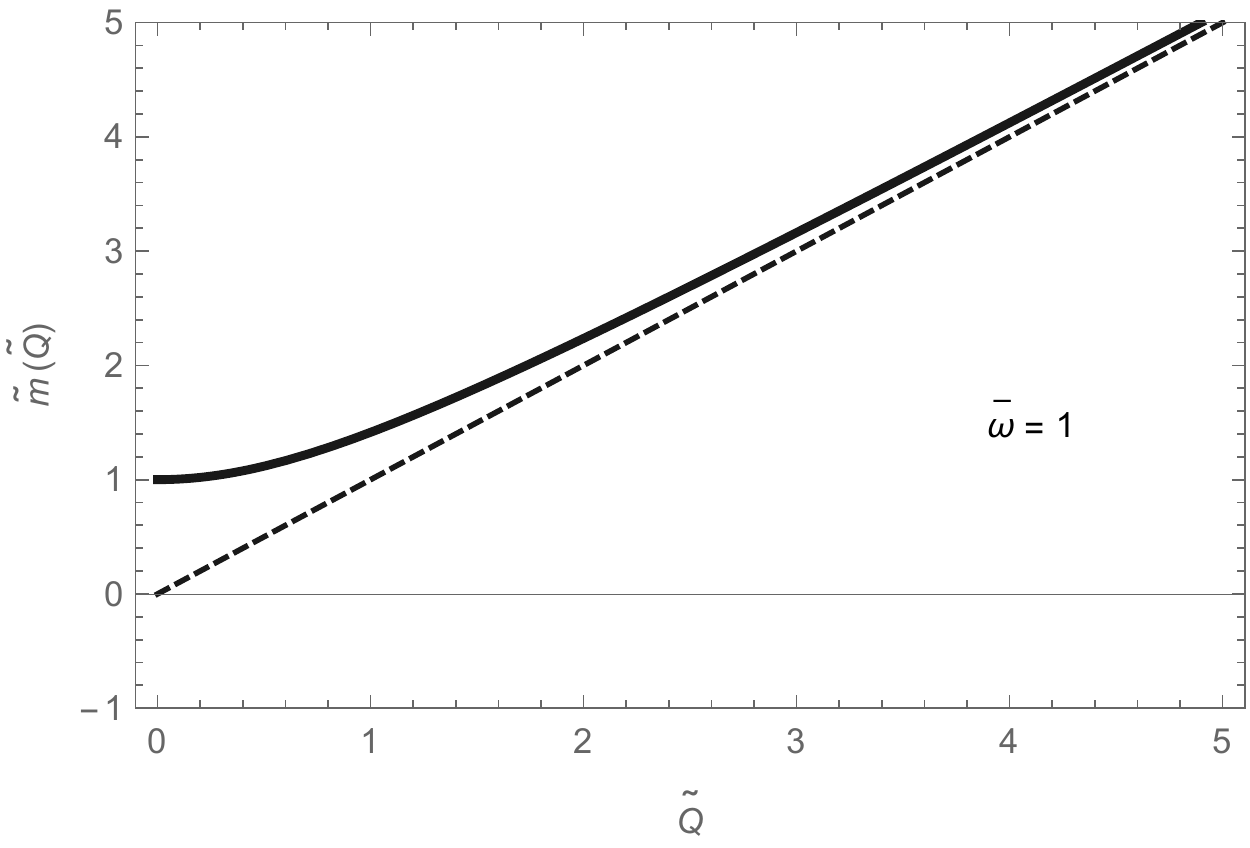}
\caption{\small{Behavior of the extremality condition for the improved Reissner-Nordstr\"{o}m spacetime near the Planck scale.}}
\label{fig:Figura2}
\end{figure}
\subsubsection{Structural stability} 
In this section we analyze the stability of the critical surfaces of the classical Reissner-Nordstr\"{o}m spacetime, defined by the solutions of the polynomial (\ref{R}). We are interested in clarifying whether the improvement leads to a drastic change in the number and form of these surfaces, or on the contrary they stay equal in number and only change smoothly. To study this aspect, we present a brief introduction to the concept of structural stability of analytic functions in one variable \cite{Tuiran, Catastrofe}.

\subsection*{Structural stability of functions in one variable}
The concept of structural stability is introduced in the framework of catastrophe theory as a basic tool in the analysis of the behavior of critical points of analytic functions, under infinitesimal variations of these functions. We say that two functions $f_{1}(r)$ y $f_{2}(r)$ are of the same type, or equivalent around $r=0$, if they have the same configuration of critical points with the same properties\footnote{For all cases this can be achieved if there exists a diffeomorphism between both functions (see \cite{Catastrofe} chapter 4).}. The analysis can be easily extended to other points $r\neq0$ of the real line by performing coordinate translations. To determine whether a function $f(r)$ is stable or not we compare it with a neighboring generic function $f_{\alpha}(r)$ given by
\begin{equation}
f_{\alpha}(r)=f(r)+\alpha(r),
\end{equation}
\noindent where $\alpha(r)$ is analytic and infinitesimally small, together with all its derivatives. We say that $f(r)$ is structurally stable at $r=0$ if it is equivalent to $f_{\alpha}(r)$ for all sufficiently small, smooth functions $\alpha(r)$.

Catastrophe Theory is concerned, among other things, about the behavior of critical points and roots of analytic functions. The concept of degeneracy is fundamental on this respect: A critical point $u$ of $f$ is called non-degenerate if the second derivative of $f$ in $u$ is different from zero. Thus, besides the condition of critical point given by
\begin{equation}
\frac{df}{dr}\Bigg|_{r=u}=0,
\end{equation}
\noindent if it's non-degenerate we also have
\begin{equation}
\frac{d^{2}f}{dr^{2}}\Bigg|_{r=u}\neq0.
\end{equation}

\noindent Equivalently, we call the function $f$ a structurally stable function at one of its critical points, if the critical point is non-degenerate. In a few words, we say concisely that the respective critical point is structurally stable \cite{Catastrofe}.

\noindent Even though the definition of structural stability is based upon properties of the critical points of a function rather than its roots, we can consider the roots as critical points of the integral of the mentioned function. We define $\bar{f}(r)$ as the integral of $f(r)$ in $r$:
\begin{equation}
\bar{f}(r)\equiv\int_{r_{0}}^{r}f(r')dr'.
\end{equation}

\noindent As a result, if $r_{c}$ is a root of $f(r)$, it is also a critical point of $\bar{f}(r)$:
\begin{equation}
\frac{d\bar{f}}{dr}\Bigg|_{r_{c}}=f(r_{c})=0.
\end{equation}
\noindent In this way, if the following condition holds
\begin{equation}
\frac{d^{2}\bar{f}}{dr^{2}}\Bigg|_{r_{c}}=\frac{df}{dr}\Bigg|_{r_{c}}\neq0,
\end{equation}
\noindent then $r_{c}$ is structurally stable. In the next section we will use this method to analyze the stability of the roots of functions such as (\ref{PCA}) or (\ref{R}), directly related to the critical surfaces.

\subsection*{Stability of the zeros of $P^{\bar{w}}_{\widetilde{Q}}$}
As described previously the equation of critical surfaces in the $d(r)=r$ approximation is given by (\ref{PCA})
\begin{equation*}
P^{\bar{w}}_{\widetilde{Q}}(\widetilde{r}) \equiv \widetilde{r}^{2}-2\widetilde{m}\widetilde{r}+\bar{w}+\widetilde{Q}^{2}=0,
\end{equation*}
\noindent which has solutions $\widetilde{r}_{\pm}^{I}=\widetilde{m}\pm\sqrt{\widetilde{m}^{2}-\bar{w}-\widetilde{Q}^{2}}$. The second derivative of $\bar{P}^{\bar{w}}_{\widetilde{Q}}$ (the primitive of $P^{\bar{w}}_{\widetilde{Q}}$) evaluated in $\widetilde{r}_{\pm}^{I}$ is given by

\begin{eqnarray}
\begin{aligned}
\frac{d^{2}\bar{P}^{\bar{w}}_{\widetilde{Q}}}{d\widetilde{r}^{2}}\Bigg|_{\widetilde{r}_{\pm}^{I}}&\equiv\frac{d{P}^{\bar{w}}_{\widetilde{Q}}}{d\widetilde{r}}\Bigg|_{\widetilde{r}_{\pm}^{I}}=2(\widetilde{r}_{\pm}^{I}-\widetilde{m}) \\
&=\pm2\sqrt{\widetilde{m}^{2}-\bar{w}-\widetilde{Q}^{2}}\neq0,
\label{ERNC}
\end{aligned}
\end{eqnarray}

\noindent which is non-zero in general, except for the quantum extremal configuration \newline $\widetilde{m}^{2}-\bar{w}-\widetilde{Q}^{2}=0$. As a consequence, it can be asserted that except for the extremal case, the critical points $\widetilde{r}_{\pm}^{I}$ are stable. This means that only a smooth displacement on critical surfaces is expected. In contrast, the mentioned extremal case, $\widetilde{r}_{+}^{I}=\widetilde{r}_{-}^{I}=\widetilde{r}_{ext}^{I}=\widetilde{m}$, leads to a null second derivative in (\ref{ERNC}), and the degenerated solution $\widetilde{r}_{ext}^{I}$ is unstable. A possible explanation of this instability is related to the disappearance of the event horizon when the discriminant $\widetilde{m}^{2}-\bar{w}-\widetilde{Q}^{2}$ changes sign from positive to negative precisely after the extremal configuration is reached.
\subsection*{Stability of the zeros of $P^{\bar{w}=0}_{\widetilde{Q}}$}
For the classical Reissner-Nordstr\"{o}m spacetime the equation (\ref{PCA}) for the horizon is reduced to the equation (\ref{R}) after setting $\bar{w}=0$
\begin{equation}
P^{0}_{\widetilde{Q}}(\widetilde{r})\equiv\widetilde{r}^{2}-2\widetilde{m}\widetilde{r}+\widetilde{Q}^{2}=0,
\end{equation}

\noindent where the solutions are given by $\widetilde{r}_{\pm}=\widetilde{m}\pm\sqrt{\widetilde{m}^{2}-\widetilde{Q}^{2}}$. The second derivative of $\bar{P}^{0}_{\widetilde{Q}}$ at each of these solutions is given by 
\begin{eqnarray}
\begin{aligned}
\frac{d^{2}\bar{P}^{0}_{\widetilde{Q}}}{d\widetilde{r}^{2}}\Bigg|_{\widetilde{r}_{\pm}}&\equiv\frac{d{P}^{0}_{\widetilde{Q}}}{d\widetilde{r}}\Bigg|_{\widetilde{r}_{\pm}}=2(\widetilde{r}_{\pm}-\widetilde{m}) \\
&=\pm2\sqrt{\widetilde{m}^{2}-\widetilde{Q}^{2}}\neq0.
\label{ERN}
\end{aligned}
\end{eqnarray}

\noindent Again the stability of the horizon solutions $\widetilde{r}_{\pm}$ is restricted to the non-extremal case $\widetilde{m}^{2}-\widetilde{Q}^{2}\neq0$; as a result, the values $\widetilde{r}_{\pm}$ change smoothly after switching $\bar{w}\neq0$ for non-extremal configurations, the quantum correction only causes a smooth change of the event horizons (internal and external) for all $\bar{w}$ infinitesimal. In contrast for the extremal case, $\widetilde{r}_{ext}=\widetilde{m}=\widetilde{Q}$, we have from (\ref{ERN})
that the second derivative is zero and the degenerate extremal solution $\widetilde{r}_{ext}$ is unstable. 
\begin{figure}[H]
\centering
\includegraphics[scale=0.8]{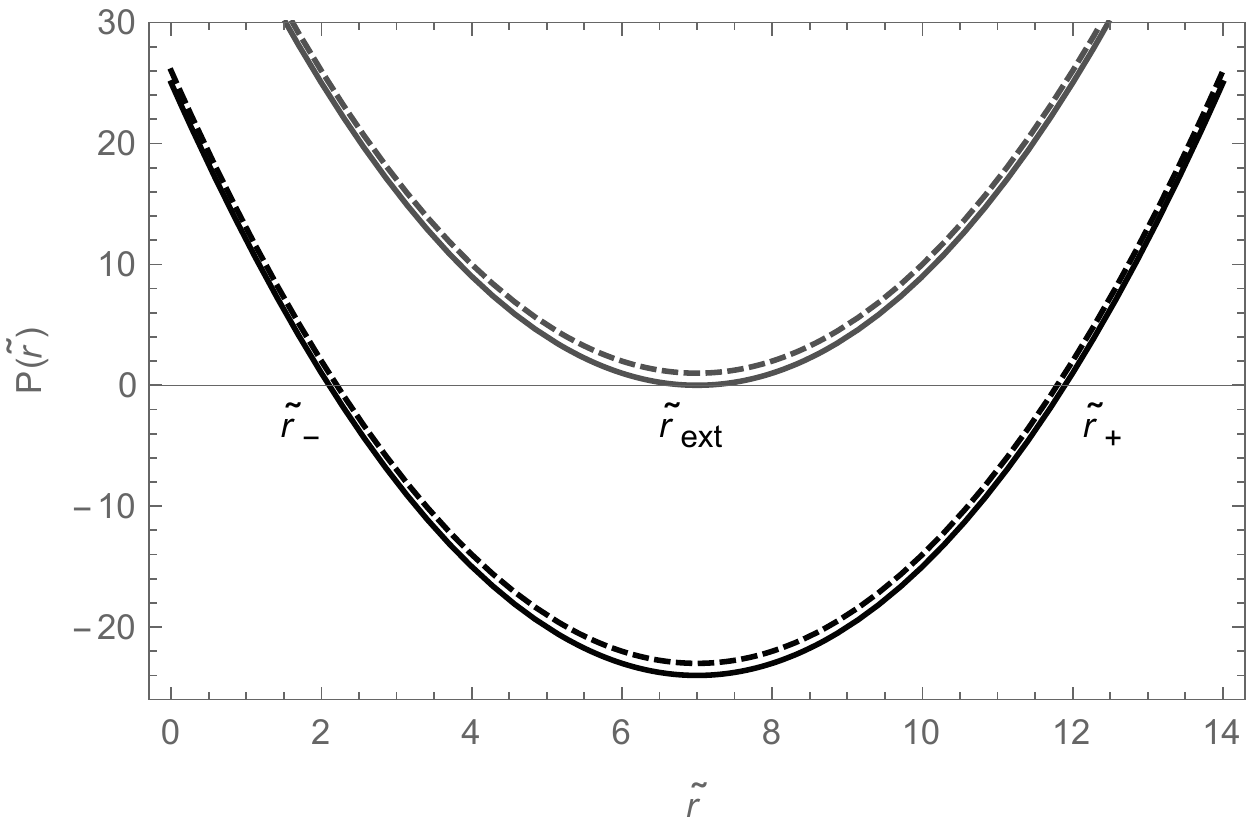}
\caption{\small{Comparison of roots of the polynomials $P^{0}_{\widetilde{Q}}(\widetilde{r})$ and  $P^{\bar{w}}_{\widetilde{Q}}(\widetilde{r})$ for non-extremal (lower ones) and extremal (upper ones) events horizons of the respectively, classical (continuous line) and improved (dashed line) Reissner-Nordstr\"{o}m space-times. The extremal dashed curve has no real roots in contrast to the extremal continuous one which has only one root ($r=\widetilde{r}_{ext}$). This can be understood due to the instability of $\widetilde{r}_{ext}$, and indicates that the improvement leads, for the extremal configuration, to the disappearance of the event horizon.}}
\label{fig:Figura3}
\end{figure}

\noindent Figure $4$ shows the dependence on $\widetilde{r}$ of the polynomials $P^{\bar{w}=0}_{\widetilde{Q}}(\widetilde{r})$ (continuous lines) and $P_{\widetilde{Q}}^{\bar{w}}(\widetilde{r})$ (dashed lines) for $\widetilde{m}=7$, $\bar{w}=1$ and several values of $\widetilde{Q}$. The roots represent the radial coordinate of the spherical event horizon surfaces in the classical (continuous) and improved (dashed) Reissner-Nordstr\"{o}m space-times respectively. As predicted by the stability of $\widetilde{r}_{\pm}$, the correction with $d(r)=r$ leads to a smooth deformation of the classical critical surfaces, as can be seen in the lower curves. The extremal configuration represents an exception where an important feature is observed, namely that $P_{\widetilde{Q}}^{\bar{w}}(\widetilde{r})$ has no real roots (upper dashed line), when $\widetilde{r}_{ext}=\tilde{m}$ is the extremal solution of $P_{\widetilde{Q}}^{\bar{w}=0}(\widetilde{r})$ (upper continuous line). This indicates the disappearance of the event horizon after turning on the quantum correction with $\bar{w}\neq0$; as predicted by the instability of $\widetilde{r}_{ext}$; the solution of $P^{\bar{w}=0}_{\widetilde{Q}}(\widetilde{r})$ with $\bar{w}=0$.      
\subsection{Thermodynamics}
\label{sec:fiddling}

\subsubsection{Area of the event horizon and surface gravity}

The area $\mathcal{A}$ of the event horizon $H$ is defined as the surface integral given by:
\begin{equation}
\mathcal{A}=\oint_{H}\sqrt{\sigma}d^{2}\theta, \label{Area de H}
\end{equation}

\noindent where $H$ is a 2-D cross-section of the event horizon surface in space-time. In order to derive several quantities at the event horizon we choose the ingoing Eddington-Finkelstein (E-F) coordinates that remove the coordinate singularity. For the improved Reissner-Nordstr\"{o}m black hole, the surface $H$ can be described in these coordinates by the conditions: $v=constant$, $r=r^{I}_{+}$, $0 \leq \theta \leq \pi$, $0 \leq \varphi \leq 2\pi$; where we have chosen the most external solution $r^{I}_{+}$ in (\ref{Imp_EH_RN}). On the other hand $\sigma$ is the determinant of $\sigma_{ab}$, the induced metric of $g_{\alpha\beta}$ on $H$, and $d^{2}\theta$ is defined as $d^{2}\theta\equiv d\theta^{1}d\theta^{2}$ with $\theta^{a}$ the angular coordinates on $H$. The induced metric on $H$ results from setting constant $r$ and $v$ in the line element:
\begin{eqnarray}
\begin{aligned}
ds^{2}_{H}&=\sigma_{ab}d\theta^{a}d\theta^{b} \\
&=g_{\theta\theta}d\theta^{2}+g_{\varphi\varphi}d\varphi^{2} \\
&=r^{2}d\theta^{2}+r^{2}\sin^2\theta d\varphi^{2}\:.
\label{ds1}
\end{aligned}
\end{eqnarray}

\noindent The determinant $\sigma$ is given by $\sigma=r^{4}\sin^{2}\theta$ and its root is $\sqrt{\sigma}=r^{2}\sin\theta$. As a result, one finds for the area in (\ref{Area de H}) the following result:

\begin{equation}
\mathcal{A}^{I}=\int^{2\pi}_{0}\int^{\pi}_{0}r^{2}\Big|_{r^{I}_{+}}\sin\theta d\theta  d\varphi =4\pi(r^{I}_{+})^{2}. \label{Area_Imp}
\end{equation}

\noindent This expression preserves the original form of the event horizon area of the classical Reissner-Nordstr\"{o}m black hole. However, the radius $r^{I}_{+}\equiv r^{I}_{+}(Q, M, \bar{w})$ depends besides on $Q$ and $M$, also on the new parameter $\bar{w}$; this dependence represents the quantum correction to the area due to the running of $G$. For the $d(r)=r$ approximation we have $r^{I}_{+}$ defined in (\ref{Imp_EH_RN_0}) and $\mathcal{A}^{I}$ takes the form:
\begin{equation}
\mathcal{A}^{I}=4\pi\left(MG_{0}+\sqrt{(MG_{0})^{2}-G_{0}\bar{w}-G_{0}Q^{2}}\right)^{2}. \label{Area_Imp2}
\end{equation}


\noindent On the other hand, the surface gravity $\kappa$; defined as the force per unit mass required by an observer at infinity to keep stationary a particle at the event horizon; can be computed, taking advantage of the property of static space-times of having the Killing vector $\boldsymbol{t}$ as tangent vector to the event horizon hypersurface. $\boldsymbol{t}$ is namely, the Killing vector corresponding to the invariance of the metric under time translations. Since the event horizon is defined as a null hypersurface, this means that $\boldsymbol{t}$, being tangent, is light-like at $r=r^{I}_{+}$, namely $\left(\boldsymbol{t}\cdot\boldsymbol{t}\right)\big|_{r^{I}_{+}}=0$; as a consequence $\boldsymbol{t}$ is simultaneously tangent and normal to the hypersurface. On the other hand the scalar function $\Phi\left(r,\theta\right) \equiv \boldsymbol{t}\cdot\boldsymbol{t}=t^{\mu}t_{\mu}$, serves to define the event horizon hypersurface, since for $r^{I}_{+}$ we have $\Phi\left(r,\theta\right)\big|_{r^{I}_{+}}=\left(t^{\mu}t_{\mu}\right)\big|_{r^{I}_{+}}=0$, fulfilling the condition of being light-like. Additionally the gradient $\partial_{\alpha}\Phi\big|_{r^{I}_{+}}$ defines a normal vector to the event horizon, this means it is proportional to $t_{\alpha} \big|_{r^{I}_{+}}$. The proportionality constant between $\partial_{\alpha}\Phi\big|_{r^{I}_{+}}$ and $t_{\alpha} \big|_{r^{I}_{+}}$ is precisely the surface gravity \cite{Wald}:  
\begin{equation}
-\partial_{\alpha}\left(t^{\beta}t_{\beta}\right)\big|_{r^{I}_{+}}=2\kappa t_{\alpha}\big|_{r^{I}_{+}} \label{GS}
\end{equation}
where we have included for convenience a factor of $2$. The components of $\boldsymbol{t}$ take the form $t_{\alpha}=(-f^{I}(r), 1, 0, 0)$ represented in the E-F coordinates (See appendix B), as a consequence evaluating $t_{\alpha}$ at $r^{I}_{+}$ where $f^{I}=0$ leads to  
\begin{equation}
t_{\alpha}\big|_{r^{I}_{+}}=(-f^{I}(r^{I}_{+}), 1, 0, 0)=(0, 1, 0, 0)=\delta^{r}_{\alpha}=\partial_{\alpha}r.
\label{tsubrH}
\end{equation}
Similarly we find for $-\partial_{\alpha}(t^{\beta}t_{\beta})$:
\begin{equation}
-\partial_{\alpha}(t^{\beta}t_{\beta})=\partial_{\alpha}f^{I}(r)=\left(\frac{df^{I}}{dr}\right)\left(\partial_{\alpha}r\right),
\label{tt}
\end{equation}
and after evaluating at $r=r^{I}_{+}$ we obtain
\begin{equation}
-\partial_{\alpha}(t^{\beta}t_{\beta})\big|_{r^{I}_{+}}=\frac{df^{I}}{dr}\Big|_{r^{I}_{+}}\left(\partial_{\alpha}r\right).
\label{ttrH}
\end{equation}
\noindent Substituting (\ref{ttrH}) and (\ref{tsubrH}) in (\ref{GS}) gives the following result for $\kappa$:
\begin{equation}
\kappa=\left(\frac{1}{2}\right)\frac{df^{I}}{dr}\Big|_{r^{I}_{+}}.
\label{GSSE}
\end{equation}
\noindent $f^{I}(r)$ is defined in (\ref{FIC}) for the improved Reissner Nordstr\"{o}m, as a consequence, after diferentiating and evaluating in $r^{I}_{+}$ we end up with the following expression\footnote{Notice that (\ref{FIC}) includes no explicit function for $G(r)$, as a result the expression we find for the surface gravity is also generic in the behavior of $G(r)$.} 
\begin{equation}
\kappa^{I}_{RN}=\frac{M\left[G(r^{I}_{+})-r^{I}_{+}G'(r^{I}_{+})\right]}{(r^{I}_{+})^{2}}-\frac{Q^{2}\left[2G(r^{I}_{+})-r^{I}_{+}G'(r^{I}_{+})\right]}{2(r^{I}_{+})^{3}},
\label{GSRNC}
\end{equation}
\noindent where we have factored terms with $M$ and terms with $Q^2$. Some important remarks related to (\ref{GSRNC}) are the following:
\begin{enumerate}
\item $\kappa^{I}_{RN}$ depends exclusively on $r$, therefore, it is constant on the event horizon; this constancy is a statement of the zeroth law of classical black hole thermodynamics, which is here generalized to the improved case.
\item $\kappa^{I}_{RN}$ goes to zero for the extremal configuration, since for this case $r^{I}_{+}=r^{I}_{-}=r^{I}_{ext}$, and the function, $f^{I}(r^{I}_{ext})$, and its derivative, $\frac{df^{I}}{dr}\Big|_{r^{I}_{+}}$, are identically zero. 

\item For $G=G_{0}=\text{constant}$, one recovers the correct result for the classical Reissner-Nordstr\"{o}m spacetime \cite{Poisson}. After applying the event horizon condition (\ref{PC}) with $\bar{w}=0$, $r^{2}-2G_{0}Mr+G_{0}Q^{2}=0$ one finds
\begin{equation}
\kappa_{class}=\frac{\left[r_{+}-MG_{0}\right]}{r^{2}_{+}}.
\end{equation}
	
\item For $Q=0$ one recovers $\kappa$ for the improved Schwarzschild black hole \cite{Tuiran}
\begin{eqnarray}
\begin{aligned}
\kappa\left[r_{+}=2MG(r_{+}), Q=0\right]&=\frac{M\left[G(r_{+})-2MG(r_{+})G'(r_{+})\right]}{4M^{2}G^{2}(r_{+})} \\ 
&=\frac{1}{4MG(r_{+})}-\frac{G'(r_{+})}{2G(r_{+})}.
\end{aligned}
\end{eqnarray}
\end{enumerate}

\subsubsection{First law of the thermodynamics of charged black holes}

The first law of the black holes mechanics for a charged and rotating black hole is given by \cite{Poisson,Bekenstein}
\begin{equation}
\delta M=\frac{\kappa}{8\pi}\delta \mathcal{A}+\Omega_{H} \delta J+\Phi_{H} \delta Q,
\label{PL}
\end{equation}
\noindent where the $J=0$ especial case, given by
\begin{equation}
\delta M-\Phi_{H}\delta Q=\left(\frac{\kappa}{8\pi}\right)\delta \mathcal{A},
\label{Primera ley}
\end{equation}

\noindent is interpreted as the first law of thermodynamics for the Reissner-Nordstr\"{o}m black hole. Here, $\kappa/2\pi$ and $A/4$ are interpreted as the temperature $T$ and the entropy $S$ of the black hole respectively. In this way expression (\ref{Primera ley}) is equivalent to
\begin{equation}
\delta M-\Phi_{H}\delta Q=T\delta S.
\label{Primera ley1}
\end{equation}

\noindent Equation (\ref{Primera ley1}) indicates that $(\delta M-\Phi_{H}\delta Q)/T$ is an exact differential, of a state function $S=S(M, Q)$. We address now the question of whether there exists a first law in the form (\ref{Primera ley1}) for the improved Reissner-Nordstr\"{o}m black hole. We start with the definition of exact differential, and its usual condition for existence \cite{Zill}.

 \begin{defn*}[\textbf{Exact Differential}] A differential expression $M(x, y)dx+N(x, y)dy$ is an \textbf{exact differential} in a region $R$ of the $xy$-plane if it corresponds to the differential of some function $f(x, y)$ defined in $R$.
 \end{defn*}
\begin{theorem*}
 Let $M(x, y)$ and $N(x, y)$ be continuous functions, with continuous first partial derivatives in a rectangular region $R$. Then a necessary and sufficient condition for $M(x, y)dx+N(x, y)dy$ to be an exact differential is that
\begin{equation*}
\frac{\partial M}{\partial y}=\frac{\partial N}{\partial x}. \label{Cruzada}
\end{equation*}

\end{theorem*}
 The states of a Reissner-Nordstr\"{o}m black hole can be labeled by the two parameters $M$ and $Q$. The corresponding state space is displayed as the Euclidean plane in two dimensions with Cartesian coordinates $x=M$, $y=Q$. Applying to the $(M,Q)$ space the previous definition of exact differential, we have that the expression $\alpha=P(M, Q)\delta M+N(M, Q)\delta Q$ is exact, if there exists a function $S$ such that $\alpha=\delta S$, that is,
\begin{equation}
P(M, Q)\delta M+N(M, Q)\delta Q=\frac{\partial S}{\partial M}\delta M+\frac{\partial S}{\partial Q}\delta Q.
\label{EDE}
\end{equation}
\noindent In consequence,
\begin{equation}
P(M, Q)=\frac{\partial S}{\partial M}, \qquad N(M, Q)=\frac{\partial S}{\partial Q}. \label{derivs_M_Q}
\end{equation}
\noindent On the other hand, if $P$ and $N$ and their first partial derivatives are continuous, then the necessary and sufficient condition for the previous theorem to be fulfilled reads
\begin{equation}
\frac{\partial P}{\partial Q}=\frac{\partial N}{\partial M}.
\label{DPM}
\end{equation}
For the improved Reissner-Nordstr\"{o}m black hole, the surface gravity (\ref{GSRNC}) can be written in the form:

\begin{equation}
k^{I}=\frac{r_{+}^{I}-M\left[r_{+}^{I}G'(r_{+}^{I})+G(r_{+}^{I})\right]}{(r_{+}^{I})^{2}}+\frac{G'(r_{+}^{I})Q^{2}}{2(r_{+}^{I})^{2}},
\label{GSRNC1}
\end{equation}
\noindent where we have substituted the general equation of the event horizon (\ref{PC_gral}). If we assume that the results (\ref{GSRNC1}) and (\ref{PERNC}) still fulfill the classical form $\delta A/(8\pi)=(1/\kappa)\delta M-(\Phi_{H}/\kappa)\delta Q$ of the first law, where $\Phi_{H}$ is the electric potential defined as
\begin{equation}  
\Phi_{H}^{I} \equiv \frac{Q}{r^{I}_{+}},
\label{PERNC}
\end{equation}
\noindent with $Q$ the electrical charge of the black hole and $r^{I}_{+}=r^{I}_{+}(M, Q)$ the radius of the external event horizon; then there must exist a scalar function $S$, such that $\delta S=(1/\kappa)\delta M-(\Phi_{H}/\kappa)\delta Q$. If $S$ exists, one could try to identify $T=\kappa^{I}/2\pi$ again with the temperature and $S$ with the entropy of the black hole, so that
\begin{equation}
\delta M-\Phi_{H}^{I}(M, Q)\delta Q\stackbin[]{?}{=}T(M, Q)\delta S(M, Q), 
\label{PLC}
\end{equation}

\noindent or explicitly
\begin{equation}
\delta M-\frac{Q}{r^{I}_{+}}\delta Q\stackbin[]{?}{=}\frac{\kappa^{I}}{2\pi}\delta S.
\label{PLC1}
\end{equation}

\noindent Replacing (\ref{GSRNC1}) in (\ref{PLC1}) and rearranging terms leads to
\begin{eqnarray}
2\pi (r_{+}^{I})^{2}\delta M-2\pi Qr_{+}^{I}\delta Q\stackbin[]{?}{=}\delta S\left\{r_{+}^{I}-M\left[r_{+}^{I}G'(r_{+}^{I})+G(r_{+}^{I})\right]+\frac{G'(r_{+}^{I})Q^{2}}{2}\right\}. \nonumber \\ 
\label{PLC2}
\end{eqnarray}
\noindent If we define
\begin{equation}
h\left(M,Q\right)\equiv\left\{r_{+}^{I}-M\left[r_{+}^{I}G'(r_{+}^{I})+G(r_{+}^{I})\right]+\frac{G'(r_{+}^{I})Q^{2}}{2}\right\}^{-1}, \label{h}
\end{equation}
\noindent then, the equation (\ref{PLC2}) can be written in the shorter form 
\begin{equation}
2\pi (r_{+}^{I})^{2}h\left(M,Q\right)\delta M-2\pi Qr_{+}^{I}h\left(M,Q\right)\delta Q\stackbin[]{?}{=}\delta S.
\label{PLC3}
\end{equation}
\noindent Where $h\left(M,Q\right)$ plays the role of an integrating factor. After comparing the previous expression with (\ref{EDE}) one can infer the following identities 
\begin{eqnarray}
P(M, Q)=2\pi(r_{+}^{I})^{2}h\left(M,Q\right) \ \text{and} \ N(M, Q)=-2\pi Qr_{+}^{I}h\left(M,Q\right). \label{D_parciales_P_y_N}
\end{eqnarray}
\noindent The fulfillment of (\ref{PLC}) implies that the partial derivatives $\partial P/\partial Q$ and $\partial N/\partial M$ must be equal. We proceed in the following lines to show this equality. From (\ref{D_parciales_P_y_N}) we see that these derivatives are given by
\begin{equation}
\frac{\partial P}{\partial Q}=4\pi hr_{+}^{I}\frac{\partial r_{+}^{I}}{\partial Q}+2\pi (r_{+}^{I})^{2}\frac{\partial h}{\partial Q}, \label{DPQ}
\end{equation}

\begin{equation}
\frac{\partial N}{\partial M}=-2\pi Qh\frac{\partial r_{+}^{I}}{\partial M}-2\pi Qr_{+}^{I}\frac{\partial h}{\partial M}.
\label{DNM}
\end{equation}
\noindent As a result, we need to calculate expressions for $\partial r_{+}^{I}/\partial Q$, $\partial r_{+}^{I}/\partial M$, $\partial h/\partial Q$ and $\partial h/\partial M$. The partial derivatives $\partial r_{+}^{I}/\partial Q$ and $\partial r_{+}^{I}/\partial M$ can be found by differentiating the event horizon's equation (\ref{PC_gral}) 
\begin{equation}
(r_{+}^{I})^{2}-2Mr_{+}^{I}G(r_{+}^{I})+G(r_{+}^{I})Q^{2}=0, 
\end{equation}
partially with respect to $Q$ and $M$ respectively. After solving for each partial derivative we find
\begin{equation}
\frac{\partial r_{+}^{I}}{\partial Q}=\frac{-2QG(r_{+}^{I})}{2r_{+}^{I}-2M\left[r_{+}^{I}G'(r_{+}^{I})+G(r_{+}^{I})\right]+Q^{2}G'(r_{+}^{I})}\:,
\label{DrQ}
\end{equation}
and
\begin{equation}
\frac{\partial r_{+}^{I}}{\partial M}=\frac{2r_{+}^{I}G(r_{+}^{I})}{2r_{+}^{I}-2M\left[r_{+}^{I}G'(r_{+}^{I})+G(r_{+}^{I})\right]+Q^{2}G'(r_{+}^{I})}\:,
\label{DrM}
\end{equation}

\noindent where we have used the implicit dependence of $G$ on $Q$ and $M$ through $r_{+}^{I}$. From (\ref{DrQ}) and (\ref{DrM}) one can see immediately that
\begin{equation}
\frac{\partial r_{+}^{I}}{\partial M}=-\frac{r_{+}^{I}}{Q}\frac{\partial r_{+}^{I}}{\partial Q}.
\label{DrM1}
\end{equation}
\noindent In a similar fashion we find for $\partial h/\partial Q$ and $\partial h/\partial M $  the following expressions
\begin{eqnarray}
\frac{\partial h}{\partial Q}=-h^{2}\left\{\left[1-M\left[2G'(r_{+}^{I})+r_{+}^{I}G''(r_{+}^{I})\right]+\frac{Q^{2}}{2}G''(r_{+}^{I})\right]\frac{\partial r_{+}^{I}}{\partial Q}+QG'(r_{+}^{I})\right\}, \ \  \label{DhQ1}
\end{eqnarray}
\begin{eqnarray}
\frac{\partial h}{\partial M}=h^{2}\frac{r_{+}^{I}}{Q}\left\{\left[1-M\left[2G'(r_{+}^{I})+r_{+}^{I}G''(r_{+}^{I})\right]+\frac{Q^{2}}{2}G''(r_{+}^{I})\right]\frac{\partial r_{+}^{I}}{\partial Q}+QG'(r_{+}^{I})\right\}+h^{2}G(r_{+}^{I}). \nonumber \\
\label{DhM}
\end{eqnarray}
\noindent It results remarkable that one can substitute (\ref{DhQ1}) in (\ref{DhM}) and finds a relation between 
$\partial h / \partial M$ and $\partial h / \partial Q$:
\begin{equation}
\frac{\partial h}{\partial M}=-\frac{r_{+}^{I}}{Q}\frac{\partial h}{\partial Q}+h^{2}G(r_{+}^{I}).
\label{DhM1}
\end{equation}
\noindent Relations (\ref{DrM1}) and (\ref{DhM1}) simplify significantly the task of verifying the equality of partial derivatives $\partial P/\partial Q$ and $\partial N/\partial M$. After replacing the mentioned relations in (\ref{DNM}) we obtain 

\begin{equation}
\frac{\partial N}{\partial M}=2\pi hr_{+}^{I}\frac{\partial r_{+}^{I}}{\partial Q}+2\pi (r_{+}^{I})^{2}\frac{\partial h}{\partial Q}-2\pi Qh^{2}r_{+}^{I}G(r_{+}^{I}),
\end{equation}
\noindent and consequently, recalling expression (\ref{DPQ}) for $\partial P / \partial Q$ leads to
\begin{equation}
\frac{\partial P}{\partial Q}-\frac{\partial N}{\partial M}=2\pi hr_{+}^{I}\frac{\partial r_{+}^{I}}{\partial Q}+2\pi Qh^{2}r_{+}^{I}G(r_{+}^{I}). \label{Eq105}
\end{equation}
\noindent By substituting (\ref{h}) in (\ref{DrQ}) the partial derivative $\partial r_{+}^{I}/\partial Q$ can be written in terms of $h$ as follows 
\begin{equation}
\frac{\partial r_{+}^{I}}{\partial Q}=-hQG(r_{+}^{I}),
\end{equation}

\noindent which finally gives 
\begin{equation}
\frac{\partial P}{\partial Q}-\frac{\partial N}{\partial M}=0,
\end{equation}
for equation (\ref{Eq105}).
\noindent The equality of partial derivatives $\partial P/\partial Q$ and $\partial N/\partial M$ indicates that the classical form of the first law could be fulfilled for the improved Reissner-Nordstr\"{o}m black hole in the form (\ref{PLC}); being the temperature $T$ directly proportional to the surface gravity $\kappa^{I}$. This result was not expected, since after the correction, the quantities $\kappa$ and $\Phi_{H}$ are modified as presented in (\ref{GSRNC1}) and (\ref{PERNC}). Also, in previous works it has been found that for improved Kerr black hole, the relationship between the Hawking temperature and surface gravity is no longer of direct proportionality, and that the original form of the first law is no longer fulfilled \cite{R-Tuiran, Tuiran}. However, for the improved Schwarzschild black hole a first law of the form $\delta S^{I}=(\delta M/\kappa)$, where the temperature is proportional to the surface gravity, has been established \cite{Bonanno-Reuter}.

\subsection{Komar Mass and Antiscreening}

We have mentioned previously that being Reissner-Nordstr\"{o}m a static spacetime, there exists a Killing vector $\boldsymbol{t}$ associated with the invariance of the metric under continuous time transformations. This permits us define a mass, taking advantage of the so-called mass Komar integral \cite{Poisson, Komar}. This integral is to be intepreted as the mass of the gravitational source:
\begin{equation}
M_{Komar}=-\frac{1}{8\pi G_{0}}\oint_{S}\nabla^{\alpha}t^{\beta}dS_{\alpha\beta}. \label{MKomar}
\end{equation}

\noindent Here $S$ is a 2-D surface at the spatial infinity. The surface element $dS_{\alpha\beta}$ is given by \cite{Poisson}
\begin{equation}
dS_{\alpha\beta}=-2n_{\left[\alpha\right.} r_{\left.\beta\right]}\sqrt{\sigma}d^{2}\theta,
\end{equation}

\noindent where $n_{\alpha}$ and $r_{\alpha}$ are respectively spacelike and timelike four-vectors normal to $S$, $\sigma$ is the determinant of $\sigma_{ab}$, the induced metric of $g_{\alpha\beta}$ on the event horizon surface $H$, and $d^{2}\theta\equiv d\theta^{1}d\theta^{2}$, where $\theta^{a}$ are the angular coordinates on $H$.

\noindent The Komar mass can be divided into two parts, one that contains only the effects of matter within the external horizon $H$, and one that is due to the distribution of matter outside of $H$ \cite{Tuiran, Poisson}. Applying Gauss' theorem we find that $M$ can be expressed as:
\begin{equation}
M=M_{H}+2\int_{\Sigma}\left(T_{\alpha\beta}-\frac{1}{2}Tg_{\alpha\beta}\right)n^{\alpha}t^{\beta}\sqrt{h}d^{3}y.
\end{equation}

\noindent $\Sigma$ is a spacelike hypersurface that extends from the event horizon to the spatial infinity, $h_{ab}$ is the induced metric in $\Sigma$ and $y^{a}$ (a=1, 2, 3) are intrinsic coordinates on the hypersurface. For the time being we are exclusively interested in the  mass $M_{H}$ contained inside the horizon, it is given by the following surface integral on $H$:
\begin{equation}
M_{H}=-\frac{1}{8\pi G_{0}}\int_{H}\nabla^{\alpha}t^{\beta}ds_{\alpha\beta},
\label{Masa en H}
\end{equation}

\noindent where
\begin{equation}
ds_{\alpha\beta}=2\xi_{\left[\alpha\right.} N_{\left.\beta\right]}\sqrt{\sigma}d^{2}\theta=(\xi_{\alpha}N_{\beta}-\xi_{\beta}N_{\alpha})\sqrt{\sigma}d^{2}\theta.
\label{ds}
\end{equation}
$\boldsymbol{\xi}$ is the most general tangent vector to the horizon, and $\boldsymbol{N}$ is an auxiliary vector that isolates the part of the metric transverse to $\boldsymbol{\xi}$. For a static, spherically symmetric space-time like Reissner-Nordstr\"{o}m and its improved version, $\boldsymbol{\xi}$ happens to be precisely $\boldsymbol{t}$, the Killing generator of the event horizon $H$ \cite{Poisson}. $\boldsymbol{N}$ fulfills the following conditions:
\begin{equation}
N_{\mu}\xi^{\mu}=N_{\mu}t^{\mu}=-1,\ N_{\mu}N^{\mu}=0. \label{Cond_Vecs_1}
\end{equation}
\noindent Substituting (\ref{ds}) in (\ref{Masa en H}) with $\xi_{\alpha}=t_{\alpha}$ leads to the following expression
\begin{equation}
M_{H}=-\frac{1}{8\pi G_{0}}\int_{H}\nabla^{\alpha}t^{\beta}(t_{\alpha}N_{\beta}-t_{\beta}N_{\alpha})ds,
\end{equation}
\noindent which can be written in the following form after rearranging indices 
\begin{equation}
M_{H}=-\frac{1}{8\pi G_{0}}\int_{H}t_{\alpha}N_{\beta}\left(\nabla^{\alpha}t^{\beta}-\nabla^{\beta}t^{\alpha}\right)ds.
\end{equation}
\noindent Applying the Killing equation for $t^{\beta}$ given by
\begin{equation}
\nabla^{\alpha}t^{\beta}+\nabla^{\beta}t^{\alpha}=0,
\end{equation}
\noindent leads to the following expression for $M_{H}$:
\begin{equation}
M_{H}=-\frac{1}{4\pi G_{0}}\int_{H}t_{\alpha}N_{\beta}(\nabla^{\alpha}t^{\beta})ds. \label{Masa en H1}
\end{equation}
Since the integral in (\ref{Masa en H1}) has to be evaluated at the (outer) event horizon, the coordinates we employ must be well behaved at $r^{I}_{+}$. Again the ingoing E-F coordinates fulfill the needed condition. As a result $\boldsymbol{t}$ is represented by $t_{\alpha}\big|_{r^{I}_{+}}=\delta_{\alpha}^r$ from (\ref{tsubrH}). Substituting it in the integrand leads to:
\begin{eqnarray}
\begin{aligned}
t_{\alpha}\big|_{r^{I}_{+}}N^{\beta}\big|_{r^{I}_{+}}(\nabla^{\alpha}t_{\beta})\big|_{r^{I}_{+}}&=\delta^{r}_{\alpha}N^{\beta}\big|_{r^{I}_{+}}(\nabla^{\alpha}t_{\beta})\big|_{r^{I}_{+}} \\
&=\left[N^{v}\big|_{r^{I}_{+}}(\nabla^{r}t_{v})\big|_{r^{I}_{+}}+N^{r}(\nabla^{r}t_{r})\big|_{r^{I}_{+}}\right].
\label{nabla t}
\end{aligned}
\end{eqnarray}

\noindent At this point we lean on the result (\ref{DCt}) from the appendix B where we have found the following expression for $\left(\nabla_{\alpha}t_{\beta}\right)\big|_{r^{I}_{+}}$:
\begin{eqnarray}
\left(\nabla_{\alpha}t_{\beta}\right)\big|_{r^{I}_{+}}=\frac{1}{2}\left. \left(\frac{\partial g_{\beta v}}{\partial x^{\alpha}}-\frac{\partial g_{\alpha v}}{\partial x^{\beta}}\right) \right|_{r^{I}_{+}}.
\end{eqnarray}
\noindent As a result the derivatives in (\ref{nabla t}) are given respectively by
\begin{equation}
\left(\nabla^{r}t_{v}\right)\big|_{r^{I}_{+}}=(g^{r\alpha}\nabla_{\alpha}t_{v})\big|_{r^{I}_{+}}=\left.\left[\frac{g^{rr}}{2}\left(\frac{\partial g_{vv}}{\partial r}\right)\right]\right|_{r^{I}_{+}}, \label{Derivada tv}
\end{equation}
\noindent and
\begin{equation}
\left(\nabla^{r}t_{r}\right)\big|_{r^{I}_{+}}=(g^{r\alpha}\nabla_{\alpha}t_{r})\big|_{r^{I}_{+}}=-\left.\left[\frac{g^{rv}}{2}\left(\frac{\partial g_{vv}}{\partial r}\right)\right]\right|_{r^{I}_{+}}, \label{Derivada tr}
\end{equation}

\noindent where only the non-zero components of the Reissner-Nordstr\"{o}m metric have been taken into account. Now substituting (\ref{Derivada tv}) and (\ref{Derivada tr}) in (\ref{nabla t}) leads to
\begin{equation}
t_{\alpha}\big|_{r^{I}_{+}}N^{\beta}\big|_{r^{I}_{+}}(\nabla^{\alpha}t_{\beta})\big|_{r^{I}_{+}}=\frac{1}{2}\left.\left[N^{v}g^{rr}\left(\frac{\partial g_{vv}}{\partial r}\right)-N^{r}g^{rv}\left(\frac{\partial g_{vv}}{\partial r}\right)\right]\right|_{r^{I}_{+}},
\label{nabla t1}
\end{equation}
\noindent and after applying\footnote{In E-F coordinates $g^{rr}=0$ everywhere, specially at the horizon. On the other hand evaluating the product $N^{\mu}t_{\mu}$ at the horizon leads to $\left.N^{\mu}t_{\mu}\right|_{r^{I}_{+}}=\left.N^{\mu}\delta^r_{\mu}\right|_{r^{I}_{+}}=\left.N^{r}\right|_{r^{I}_{+}}$ which turns out to be equal to $-1$ from condition $N^{\mu}t_{\mu}=-1$ in (\ref{Cond_Vecs_1}). The rest of components of the auxiliary vector $\boldsymbol{N}$ can be set to zero without losing generality.} $\left. g^{rr}\right|_{r^{I}_{+}}=0$ and $\left. N^{r}\right|_{r^{I}_{+}}=-1$ in (\ref{nabla t1}), the integral (\ref{Masa en H1}) turns out to be
\begin{equation}
M_{H}=-\frac{1}{8\pi G_{0}}\int_{H}\left(\frac{\partial g_{vv}}{\partial r}\right)g^{rv}ds. \label{Masa en H2}
\end{equation}

\noindent Evaluating the derivative $\partial g_{vv} / \partial r$ at the outer horizon $r^{I}_{+}$ gives the following expression
\begin{eqnarray}
\begin{aligned}
\left(\frac{\partial g_{vv}}{\partial r}\right)\bigg|_{r^{I}_{+}}&=-\frac{\partial}{\partial r}\left(1-\frac{2G(r)M}{r}+\frac{G(r)Q^2}{r^{2}}\right)\bigg|_{r^{I}_{+}} \\
&=\frac{2M}{r^{2}}\left[G'(r)r-G(r)\right]\bigg|_{r^{I}_{+}}-\frac{Q^{2}}{r^{3}}\left[G'(r)r-2G(r)\right]\bigg|_{r^{I}_{+}} \label{deriv1}.
\end{aligned}
\end{eqnarray}

\noindent On the other hand the area element is defined as $ds\left|_{r^{I}_{+}}\right.=(r^{I}_{+})^{2}\sin\theta d\theta  d\varphi$. Substituting this term and (\ref{deriv1}) in (\ref{Masa en H2}) gives finally:
\begin{eqnarray}
M_{H}&=&-\frac{1}{8\pi G_{0}}\int^{2\pi}_{0}\int^{\pi}_{0}(r^{I}_{+})^{2}\sin\theta d\theta  d\varphi \times \nonumber \\ &\times& \left\{\frac{2M}{r^{2}}\left[G'(r)r-G(r)\right]\bigg|_{r^{I}_{+}}-\frac{Q^{2}}{r^{3}}\left[G'(r)r-2G(r)\right]\bigg|_{r^{I}_{+}}\right\} \nonumber\\
&=&-\frac{M}{G_{0}}\left[G'(r^{I}_{+})r^{I}_{+}-G(r^{I}_{+})\right]+\frac{Q^{2}}{2G_{0}r^{I}_{+}}\left[G'(r^{I}_{+})r^{I}_{+}-2G(r^{I}_{+})\right]. \nonumber \\ \label{M_h_1}
\end{eqnarray}
Expression (\ref{M_h_1}) resembles our previous result (\ref{GSRNC}) for $\kappa^{I}_{RN}$. In fact a straightforward multiplication leads to the following relation
\begin{eqnarray}
M_{H}=\frac{\left(r^{I}_+\right)^2}{G_0}\kappa^{I}_{RN}
,
\end{eqnarray}
which turns out to be the special case of the generalized Smarr formula for a charged black hole when the angular momentum $J_{H}$ is zero \cite{Poisson,Smarr}:
\begin{eqnarray}
M_{H}=\left(\frac{1}{4\pi G_0}\right)\mathcal{A}^{I}\kappa^{I}_{RN}, \label {Smarr RNs}
\end{eqnarray}
where we have applied the previously found expression $\mathcal{A}=4\pi(r^{I}_{+})^{2}$ in (\ref{Area_Imp}) for the area of the event horizon\footnote{Notice that both $M_H$ and $\kappa^{I}_{RN}$ have contributions terms with $Q^2$, never the less expression (\ref{Smarr RNs}) has no explicit terms in $Q$. This is the case for the so-called generalized Smarr formula, which holds for the classical Kerr-Newman solution for rotating charged black holes. An alternative expression with $M$ and $J$ instead of $M_{H}$ and $J_{H}$, including an explicit term with $Q$, is also available in the literature (See \cite{Poisson}, for example).}. \\

In order to analyze the long range effect on $M_{H}$ for $r\to \infty$ of the RG improvement, we express equation (\ref{M_h_1}) in terms of the explicit expression (\ref{G4}) of $G(r)$ for the $d(r)=r$ approximation. The derivative of $G(r)$ evaluated at $r^{I}_{+}$ turns out to be
\begin{equation}
G^{'}(r^{I}_{+})=\frac{2G^{2}_{0}r^{I}_{+}\bar{w}}{\left[(r^{I}_{+})^{2}+\bar{w}G_{0}\right]^{2}}. \label{Derivada de G}
\end{equation}

\noindent Substituting (\ref{Derivada de G}) and (\ref{G4}) in (\ref{M_h_1}) leads to the expression:
\begin{equation}
M_{H}=M\frac{(r^{I}_{+})^{2}\left[(r^{I}_{+})^{2}-\bar{w}G_{0}\right]}{\left[(r^{I}_{+})^{2}+\bar{w}G_{0}\right]^{2}}-\frac{Q^{2}(r^{I}_{+})^{3}}{\left[(r^{I}_{+})^{2}+\bar{w}G_{0}\right]^{2}}. \label{M de H con w}
\end{equation}

\begin{figure}[H]
\centering
\includegraphics[scale=0.48]{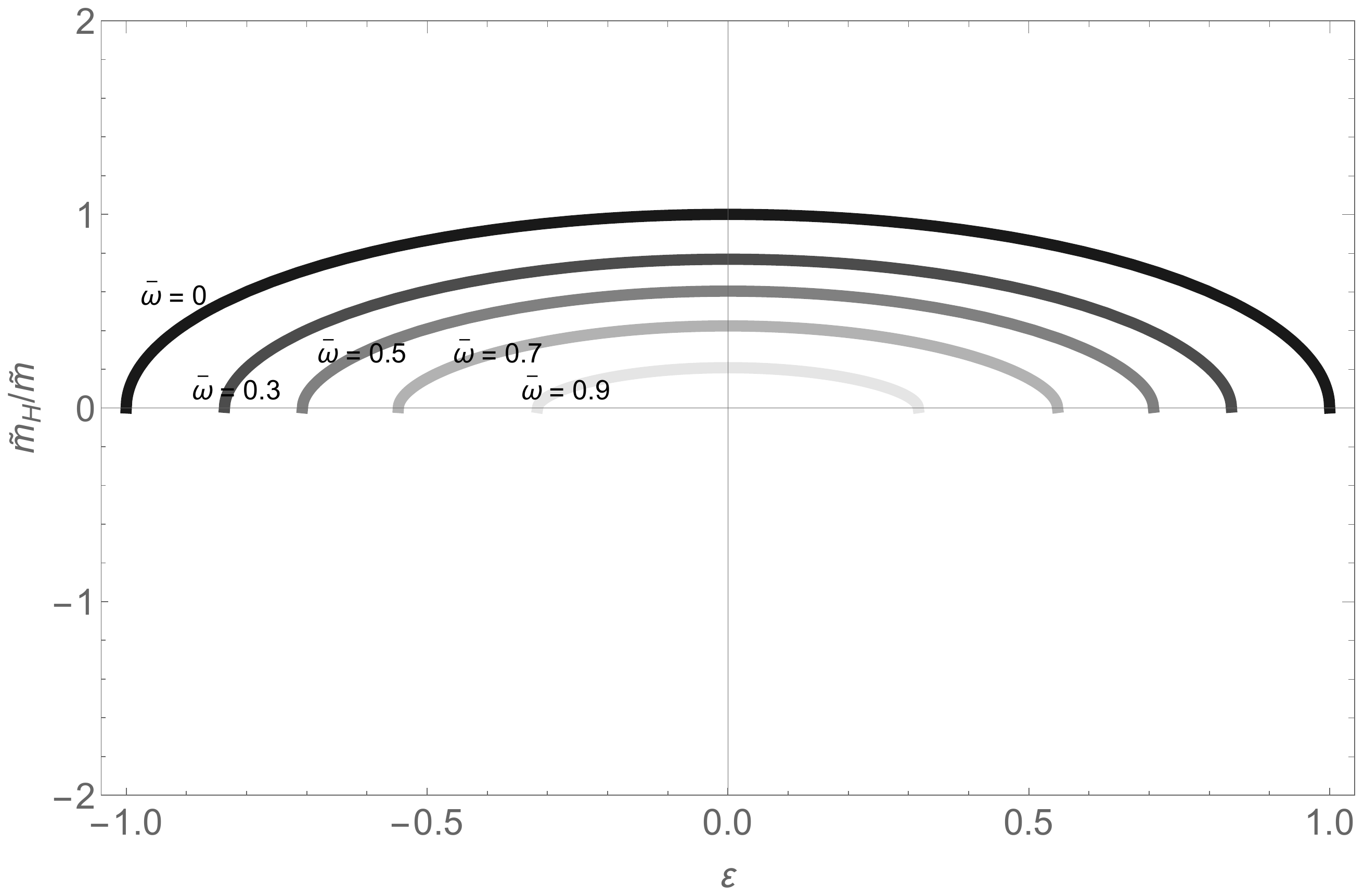}
\caption{\small{$\frac{\widetilde{m}_{H}}{\tilde{m}}$} vs $\varepsilon$ curves with increasing values from $\bar{w}=0$ (black) to $\bar{w}=0.9$ (white).}
\label{Fig 5}
\end{figure}

\noindent After applying the definitions presented in appendix A, a dimensionless form of the equation (\ref{M de H con w}) can be deduced:

\begin{eqnarray}
\frac{\widetilde{m}_{H}}{\widetilde{m}}=\frac{(\widetilde{r}_{+}^{I})^{2}}{\left( (\widetilde{r}_{+}^{I})^{2}+\bar{w}\right) ^{2}}%
\left\{ (\widetilde{r}_{+}^{I})^{2}-\bar{w} -\left(\frac{\widetilde{r}_{+}^{I}}{\widetilde{m}}\right)\tilde{Q}%
^{2}\right\}, \label{m de H con w}
\end{eqnarray}
\noindent where $\widetilde{r}_{+}^{I}$ is given in (\ref{RRNC}). Substituting the charge-to-mass ratio $\varepsilon \equiv \tilde{Q}/\tilde{m}$ into (\ref{RRNC}) and (\ref{m de H con w}) leads to the following expressions: 

\begin{equation}
\frac{\widetilde{r}_{+}^{I}}{\tilde{m}}=1+\sqrt{1-\varepsilon ^{2}-\frac{\bar{w}}{%
\tilde{m}^{2}}}\:,
\end{equation}%
\begin{equation}
\frac{\tilde{m}_{H}}{\tilde{m}}=\frac{\left(\frac{\widetilde{r}_{+}^{I}}{\tilde{m}}\right)^2}{%
\left[\left(\frac{\widetilde{r}_{+}^{I}}{\tilde{m}}\right)^2+\frac{\bar{w}}{\tilde{m}^{2}}%
\right] ^{2}}\left\{ \left(\frac{\widetilde{r}_{+}^{I}}{\tilde{m}}\right)^2-\frac{\bar{w}}{%
\tilde{m}^{2}}-\left(\frac{\widetilde{r}_{+}^{I}}{\tilde{m}}\right)\varepsilon ^{2}\right\}, 
\end{equation}%
which for the $\bar{w}=0$ classical Reissner-Nordstr\"{o}m spacetime
become exclusively $\varepsilon $-dependent:%
\begin{eqnarray}
\frac{\tilde{r}_+}{\tilde{m}}=\left( 1+\sqrt{1-\varepsilon ^{2}}\right), \\
\frac{\tilde{m}_{H}}{\tilde{m}}=1-\frac{\varepsilon ^{2}}{\left( 1+
\sqrt{1-\varepsilon ^{2}}\right) }.
\end{eqnarray}
\noindent Figure \ref{Fig 5} shows for several curves with increasing values of $\bar{w}$ from black to white; the dependence of $\widetilde{m}_{H}/\tilde{m}$ on $\varepsilon$ for the whole domain \newline $-1\leq\varepsilon\leq1$ of values that preserve the (classical) event horizon. The fraction of masses is always lower than $1$ and reaches $1$ only for $\bar{w}=0$, $\varepsilon=0$ (Schwarzschild spacetime) at the top of the black outermost curve. This ``lack" of mass at the event horizon indicates a contribution to the mass $\tilde{m}$ at infinity due to the gravitational field outside the black hole. The lighter gray curves for $\bar{w}\neq0$ show both the narrowing of domains of $\varepsilon$ that still permit the existence of an event horizon, and the decrease of $\widetilde{m}_{H}/\tilde{m}$ not even reaching $1$ for $\varepsilon=0$.   
\noindent Figure 6 shows a coincident behavior for \small{$\frac{\widetilde{m}_{H}}{\tilde{m}}$} vs \small{$\frac{\widetilde{r}_{+}^{I}}{\tilde{m}}$}. Each curve has been computed running values of $\varepsilon$ from $-1$ to $1$. The range of existence of an event horizon for the improved Reissner-Nordstr\"{o}m spacetime decreases (from black to white) with increasing $\bar{w}$, as can be seen by the ordered shortening of length of the curves from top (black) to bottom (light gray). The values of \small{$\frac{\widetilde{m}_{H}}{\tilde{m}}$} are always lower than $1$ for $\bar{w}\neq 0$ and reach $1$ for the classical case at $\bar{w}=0$ (the longer black curve on top of the others). All curves coincide at \small{$\frac{\widetilde{m}_{H}}{\tilde{m}}=0$} (complete screening at event horizon) for the extremal black hole with $1-\varepsilon ^{2}-\frac{\bar{w}}{\tilde{m}^{2}}=0$.     
\begin{figure}[H]
\centering
\includegraphics[scale=0.38]{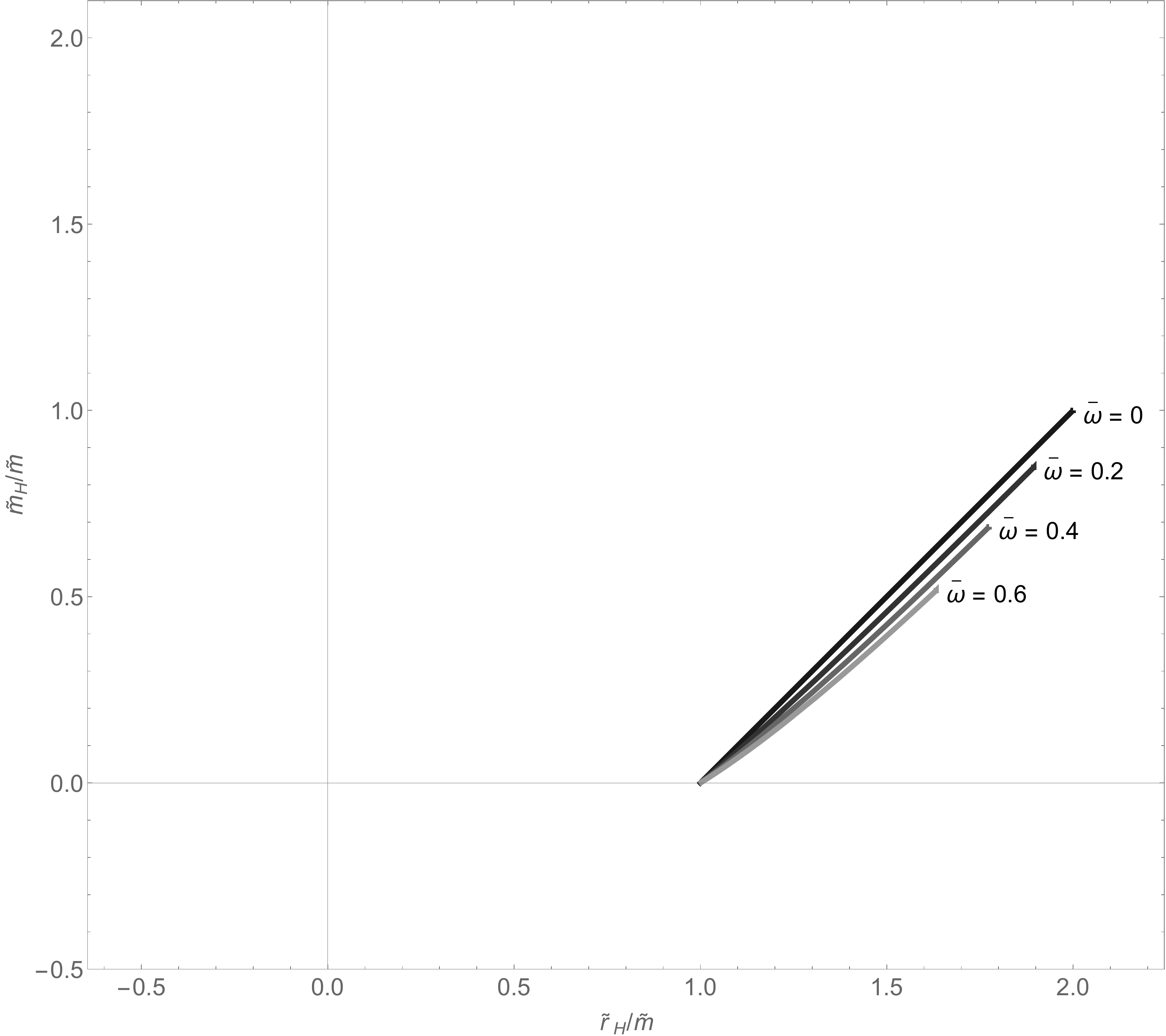}
\caption{\small{$\frac{\widetilde{m}_{H}}{\tilde{m}}$} vs $\frac{\widetilde{r}_{+}^{I}}{\tilde{m}}$ curves with increasing values from $\bar{w}=0$ (black) to $\bar{w}=0.9$ (white).}
\label{Fig 6}
\end{figure}
\noindent In figure 6 we have explored the values of $\frac{\widetilde{m}_{H}}{\tilde{m}}$, finding that except for the classical case ($\bar{w}=0$), this fraction is lower than 1 for fixed $\tilde{m}$ and all possible values of $\varepsilon$ and $\bar{w}$ that preserve the event horizon of the improved Reissner-Nordstr\"{o}m spacetime. This behavior shows a ``lack" of mass that must be recovered at infinity since $\tilde{m}$ amounts to the complete mass of the black hole measured by distant observers. We thus interpret this behavior as the result of an antiscreening effect that can only be due to the gravitational field outside the event horizon.

\newpage
\section{Discussion and Conclusions}
\noindent The results of this work can be summarized as follows: \\

\noindent The RG-improvement of the Reissner-Nordstr\"{o}m spacetime through the running Newton constant, leads to a smooth displacement of the usual critical surfaces and to the preservation of their amount for smooth variations of $\bar{w}$. This is a consequence of the local stability of solutions $\widetilde{r}_{\pm}$ of the classical spacetime, except for the extremal case where $\tilde{m}^2=\tilde{Q}^2$ (See figure 4). A Further investigation of the extremal configuration an the possibility of existence of ``quantum naked singularities" goes beyond the scope of our current approach. The process of improvement and subsequent cutoff identification will surely need a finer formulation where more degrees of freedom and coupling constants are taken into account.
\newline \\
An extremal configuration related to the parameter $\bar{w}$ arises as a consequence of the RG-improvement in accordance with previous results \cite{Bonanno-Reuter,R-Tuiran}. This configuration is reached at the Planck scale, where $\widetilde{m}\to\sqrt{\bar{w}}$ with $\widetilde{Q}$ $\to 0$, and it must be distinguished from the extremal state at $\widetilde{m} = \widetilde{Q}$ for the classical Reissner-Nordstr\"{o}m spacetime, where $\widetilde{Q}$, the macroscopic charge of the black hole remains finite. The similarity of figure $3$ for $\widetilde{m}(\widetilde{Q})$ in our present work with figure $6$ for $\widetilde{m}(\widetilde{a})$ in reference \cite{R-Tuiran} is suggestive. In reference \cite{Bonanno-Reuter} the features of a possible remaining state at the Planck scale are explored, where among several results, $\widetilde{m}$ tends to a final and unique value $\sqrt{\bar{w}}$ and the temperature reaches the zero value after an infinite time interval as prescribed by  the third law of black hole thermodynamics. Also the connection to the Reissner-Nordstr\"om spacetime for the mentioned extremal state at the quantum level is investigated. The result of our present work for which $\widetilde{m}$ $\to$ $\sqrt{\bar{w}}$ for $\widetilde{Q}\to 0 $ joined with the result previously found in \cite{R-Tuiran} for which  also $\widetilde{m}$ $\to$ $\sqrt{\bar{w}}$ for $\widetilde{a}\to 0$ contribute to add hints into the possibility of existence of a ``cold'' Planck-sized remnant which would solve the loss of information problem in the process of evaporation of black holes \cite{Bonanno-Reuter,Markov}. \newline

\noindent After finding in (\ref{GSRNC}) an exact expression for the surface gravity $\kappa^{I}$ at the event horizon of the improved Reissner-Nordstr\"om spacetime, we have shown the exactness of the following differential form:  
\begin{equation}
\delta S\equiv \left(\frac{1}{\kappa^{I}}\right)\delta M-\left(\frac{1}{\kappa^{I}}\right) \left(\frac{Q}{r^{I}_{+}}\right)\delta Q \label{1st Law}
\end{equation}
This result indicates that there exists a function $S\left(M,Q\right)$  for which the surface gravity serves as integrating factor, and that an appropriate version of the first law of black hole thermodynamics exists for the improved Reissner-Nordstr\"{o}m spacetime at least within the framework of ``phenomenological" thermodynamics. The interpretation of $S$ in (\ref{1st Law}) as an entropy; its connection to the corrected area (\ref{Area_Imp}), and a further comparison with equivalent results including microscopic degrees of freedom and an underlying statistical mechanics should be the subject of future investigations.\\     

\noindent We have calculated the Komar integral for the mass at the event horizon. The result in (\ref{M_h_1}) is a generic expression independent of the explicit form of $G(r)$ and it becomes:
\begin{equation}
    M_H=M-\frac{Q^{2}}{r_{+}} \label{M_h_RNs_CLass}
\end{equation}
 for $G(r)=G_0$, the classical case. The $Q^2$ term in (\ref{M_h_RNs_CLass}) stands for a mass-screening effect due to pure electrostatic energy at the event horizon, and it is interpreted as the energy necessary to bring together the total charge $Q$ from infinity to $r_{+}$: It is subtracted since it has to be expended against electrostatic repulsion, in order to create the source with mass $M$ and charge $Q$. Equation (\ref{M_h_RNs_CLass}) is valid even for $r\neq r_{+}$ as has been shown in \cite{Cohen1,Cohen2,Israel} and references therein. It has been interpreted as an effective gravitational mass $M(r)=M-Q^2/r$ due to the electric field which permeates all the space \cite{Cohen2} and predicts interesting non-trivial effects as gravitational bounce and radial test-particle oscillations which are still subject of investigation \cite{Cohen1,Israel,Gron,Kadir1,Kadir2}.\newline 
 \noindent We are interested in analyzing the long range effect of the running $G(r)$ on $M_{H}$ for $r\to \infty$, and check further deviations at the event horizon from $M$, the Komar mass at infinity, on top of the (classical) electrostatic ones in (\ref{M_h_RNs_CLass}). On this account we have included in (\ref{M_h_1}) the explicit expressions (\ref{G4}) and (\ref{Derivada de G}) for the $d(r)=r$ approximation. The result is expressed in equation (\ref{M de H con w}), whose expansion to the first order in $\frac{\bar{w}G_{0}}{r^{2}}$ is given by:
\begin{eqnarray}
M_{H} &=&M\left( 1-3\frac{\bar{w}G_{0}}{(r^I_+)^{2}}\right) -\frac{Q^{2}}{r^I_+} \label{1er orden Komar}
\end{eqnarray}
\noindent In the large-distance regime the leading quantum gravitational effects are expected to be fully encoded in the running Newton's constant \cite{Reuter_1}. This is coherent with a microscopic picture of a gravitational vacuum populated by virtual graviton pairs being attracted towards the ``undressed" mass $M_H$ at the event horizon and leading to a positive mass dressing for $r>r_+$ and a final ``dressed" mass $M>M_H$ at infinity. It is clear from (\ref{1er orden Komar}) and the previous analysis of figures 5 and 6, that the additional terms in $\bar{w}$ for the long-range regime amount an additional subtraction to $M$ at $r^I_+$ for all possible configurations in the space of parameters $(M,Q,\bar{w})$ avoiding naked singularities. We hypothesize that the mass recovered at infinity is a result of the above explained gravitational antiscreening effect. The question of how specifically this antiscreening is carried out and to what extent it relates to the electrostatic term is outside the aims of our current work. It is never the less an important matter for future research.  \newline      

As concluding remarks, we find appropriate to contrast our methodology and results with recent works concerning quantum improved versions of the Reissner-Nordstr\"om spacetime \cite{Koch-Gonz, Ishibashi}. They are based upon respective results of Daum, Harst and Reuter \cite{DHR1,DHR2,Harst} and Eichhorn and Versteegen \cite{Eich-Verst} about the running gauge couplings in asymptotically safe quantum gravity without and with matter content. The inclusion of the running $U(1)$ coupling or fine structure constant $\alpha$ in addition to the running Newton's constant is a common feature of these studies. We have conversely restricted our analysis to the effects of the running Newton's constant because of two main reasons. First, we have concentrated great deal of our efforts in investigating the effects of quantum gravity in the long range regime ($r\to\infty$, $k \to 0$); in contrast the running gauge couplings are expected to exert great influence at short distances and large (GUT, Planckian) energy scales ($r\to 0$, $k \to \infty$).  In the second place, we have chosen, instead of coping with the added action of several variables like $G$, $\Lambda$ and $\alpha$ to isolate the effects of only one. Some of the results of references \cite{Koch-Gonz} and \cite{Ishibashi} endorse our approach for the long range regime like the existence of smoothly shifted horizons and the ``loss" of mass at the event horizon. Our result of integrability of the $dS$ form in (\ref{1st Law}) is nevertheless valid for any generic function $G(r)$, being it related to long or short range regimes. 

\section*{Acknowledgements}
The authors wish to thank Prof. Martin Reuter from Mainz Institute of Physics for stimulating discussions and ideas about this work. The second author thanks the German Service of Academic Exchange (DAAD) for its constant support.
\section*{Appendix A \\ \\ \textbf{Planck Units and Dimensionless Quantities}} \label{App A}

\noindent In this appendix we define some useful dimensionless variables, employed throughout this work. As a notation we use the tilde 
$\ \widetilde{}\ $ in order to express the dimensionless character of a quantity.

Starting with the following fundamental constants
\begin{eqnarray*}
G_{0} &=&6.67259\times 10^{-11}\ m^{3} kg^{-1} s^{-2} \\
c &=&2.99792458\times 10^{8}\ m\ s^{-1} \\
\hbar &=&1.05457266\times 10^{-34}\ J\ s
\label{CF}
\end{eqnarray*}

\noindent One can construct the following units that define the Planck scale: 
\begin{eqnarray}
l_{p} &=&\sqrt{\frac{\hbar G_{0}}{c^{3}}}=1.61605\times 10^{-35}\ m \\
m_{p} &=&\sqrt{\frac{\hbar c}{G_{0}}}=2.17671\times 10^{-8}\ kg \\
t_{p} &=&\sqrt{\frac{\hbar G_{0}}{c^{5}}}=5.39056\times 10^{-44}\ s \\
q_{p} &=&\sqrt{4\pi\epsilon_{0}\hbar c}=1.87555\times 10^{-18}\ C
\end{eqnarray}

\noindent In the gaussian units system $4\pi\epsilon_{0}=1$ there fore $q_{p}=\sqrt{\hbar c}$. By setting $\hbar=c=1$ one obtains for the Planck length, mass, time and charge, respectively,  
\begin{eqnarray}
l_{p}=\sqrt{G_{0}}\;,\;m_{p}=\frac{1}{\sqrt{G_{0}}}\;,\;t_{p}=\sqrt{G_{0}}\;,\;q_{p}=1
\label{lmt Planck}
\end{eqnarray}

\noindent With the expressions in (\ref{lmt Planck}) one can define the following dimensionless quantities:
\begin{eqnarray}
\tilde{r}=\frac{r}{l_{p}}=\frac{r}{\sqrt{G_{0}}}\;,\;\tilde{M}=\frac{M}{m_{p}}=\sqrt{G_{0}}M\;,\;\tilde{Q}=\frac{Q}{q_{p}}=Q  \notag 
\end{eqnarray} 

\noindent As a result the radial coordinate $r$, the black hole's mass $M$ and its geometric mass $m=MG_0$, have their respective dimensionless quantities $\tilde{r}$, $\tilde{M}$, and $\tilde{m}$ as follows:
\begin{equation}
r=\tilde{r}\sqrt{G_{0}}\;,\;M=\frac{\tilde{M}}{\sqrt{G_{0}}}\;,\;m=MG_{0}=\sqrt{G_{0}}\tilde{M}\ ,\ m=\tilde{m}\sqrt{G_0}  
\label{CD}
\end{equation}

\section*{Appendix B \\ \\ \textbf{The improved Reissner-Nordstr\"{o}m space-time in Eddington-Finkelstein coordinates}} \label{App B}

\noindent A spherically symmetric and static space-time represented in the Schwarzschild coordinates $(t,r,\theta,\varphi)$ is given in (\ref{RN_1})
\begin{equation}
ds^{2}=-f(r)dt^{2}+f(r)^{-1}dr^{2}+r^{2}d\Omega^{2}.
\label{MSE}
\end{equation}
\noindent Provided that these coordinates aren't regular at the event horizon, it turns out necessary to choose a regular representation that permits the computation of quantities like the surface gravity or the Komar mass. The Eddington-Finkelstein coordinates (E-F) fulfill this requirement; they result from subtracting (outgoing) or adding (ingoing)  the singular fraction at the event horizon: 
\begin{equation}
du=dt-dr^{*} \text{ (outgoing)}, \label{EFS}
\end{equation}
\begin{equation}
dv=dt+dr^{*} \text{ (ingoing)}, \label{EFE}
\end{equation}
\noindent where $dr^{\ast}$ is defined as
\begin{equation}
dr^{*}=\frac{dr}{f(r)}.
\end{equation}
\noindent Choosing the ingoing transformation (\ref{EFE}) leads to the following squared length: 
\begin{equation}
ds^{2}=-f(r)dv^{2}+2dvdr+r^{2}d\Omega^{2}.
\label{MEFE}
\end{equation}
\noindent Alternatively the outgoing choice renders the following result: 
\begin{equation}
ds^{2}=-f(r)du^{2}-2dudr+r^{2}d\Omega^{2}.
\label{MEFS}
\end{equation}
\noindent Both expressions (\ref{MEFE}) and (\ref{MEFS}) lead to well behaved representations at the event horizon where $f(r_{H})=0$. \\
\noindent In order to represent the improved Reissner-Nordstr\"{o}m space-time in the ingoing E-F coordinates we substitute $f_{I}(r)$ from (\ref{FIC}) in (\ref{MEFE}), thus obtaining:
\begin{equation}
ds^2=-\left(1-\frac{2G(r)M}{r}+\frac{G(r)Q^{2}}{r^{2}}\right)dv^2+2drdv+r^2d\theta^2+r^2\sin^2\theta d\varphi^2. 
\label{MRNCEFE}
\end{equation}
\subsection*{Some useful expressions in E-F coordinates}
\noindent Starting from the fact that the metric (\ref{MSE}) is $t$-independent 
\begin{equation}
\frac{\partial g_{\mu\nu}}{\partial t}=0,
\end{equation}
we define the temporal Killing vector $\mathbf{t}$ as
\begin{equation}
\boldsymbol{t}=\frac{\partial}{\partial t},
\end{equation} 
as the partial derivative in the temporal coordinate.
\noindent In the ingoing E-F coordinates  $x^{\mu}=(v,r,\theta,\varphi)$ we find for $\boldsymbol{t}$ 
 \begin{equation}
t^{\beta}=\frac{\partial x^{\beta}}{\partial v}=\left(\frac{\partial v}{\partial v},\frac{\partial r}{\partial v},\frac{\partial \theta}{\partial v},\frac{\partial \varphi}{\partial v}\right)=\delta^{\beta}_{v}.
\label{tsup}
\end{equation}

\noindent Lowering the index: 
\begin{equation}
t_{\beta}=g_{\beta\sigma}t^{\sigma}=g_{\beta\sigma}\delta^{\sigma}_{v}=g_{\beta v}. \label{gbetav}
\end{equation}
\noindent Substituting the metric components (\ref{MEFE}) in (\ref{gbetav}) gives 
\begin{equation}
t_{\beta}=g_{\beta v}=(g_{vv}, g_{rv}, g_{\theta v}, g_{\varphi v})=(-f(r), 1, 0, 0).
\label{tsub}
\end{equation}

\noindent We will use the last expression in order to compute the surface gravity at the event horizon of the improved Reissner-Nordstr\"{o}m space-time. \\

\noindent Another important expression to be employed in the derivation of the Komar mass is the covariant derivative $\nabla_{\nu}t_{\mu}$:
\begin{eqnarray*}
\nabla_{\nu}t_{\mu}=\nabla_{\nu}(g_{\mu\alpha}t^{\alpha})=g_{\mu\alpha}\nabla_{\nu}(t^{\alpha})+t^{\alpha}\nabla_{\nu}(g_{\mu\alpha}) \noindent \\
=g_{\mu\alpha}\left(\frac{\partial t^{\alpha}}{\partial x^{\nu}}+\Gamma_{\nu\sigma}^{\alpha}t ^{\sigma}\right) 
=g_{\mu\alpha}\left[\frac{\partial\left(\delta_{v}^{\alpha}\right)}{\partial x^{\nu }}+\Gamma_{\nu \sigma}^{\alpha}\delta_{v}^{\sigma}\right]
=\Gamma_{\mu\nu v};
\end{eqnarray*} applying the usual definition of the Christoffel symbol $\Gamma_{\alpha\beta\gamma}$ to the previous expression leads to
\begin{eqnarray}
\nabla_{\nu}t_{\mu}=\frac{1}{2}\left(\frac{\partial g_{\mu v}}{\partial x^{\nu }}+\frac{\partial g_{\mu \nu }}{\partial v}-\frac{\partial g_{\nu v}}{\partial x^{\mu}}\right)
=\frac{1}{2}\left( \frac{\partial g_{\mu v}}{\partial x^{\nu }}-\frac{\partial g_{\nu v}}{\partial x^{\mu}}\right), \label{DCt}
\end{eqnarray}
where we have used the cyclic character of the variable $v$ in the components $g_{\mu\nu}$. An important fact is that expression (\ref{DCt}) fulfills the Killing equation $\nabla_{\nu}t_{\mu}+\nabla_{\mu}t_{\nu}=0 $ as expected. The final expression is the one applied in (\ref{nabla t}), in order to find the mass $M_{H}$ inside the event horizon of the Reissner-Nordstr\"{o}m black hole.

\bibliographystyle{model1-num-names}

\end{document}